\documentclass[useAMS,usenatbib]{mn2e}
\usepackage{graphicx}
\usepackage{dcolumn}
\usepackage{bm}
\usepackage{amssymb, amsmath}
\usepackage{color}
\usepackage{comment}
\usepackage{aas_macros}
\usepackage{hyperref}

\newcommand\beq{\begin{equation}}
\newcommand\eeq{\end{equation}}

\usepackage{mn2e-bst}

\begin{document}

\title{Characterizing Extragalactic Anomalous Microwave Emission in NGC\,6946 with CARMA}

\author[Brandon Hensley, Eric Murphy, and Johannes Staguhn]{Brandon Hensley$^1$\thanks{E-mail:bhensley@princeton.edu}, Eric Murphy$^2$, \& Johannes Staguhn$^{3,4}$ \\
$^1$Department of Astrophysical Sciences, Princeton
University, Princeton, NJ 08544, USA \\
$^2$Spitzer Science Center, California Institute of Technology, 1200 East California Boulevard, Pasadena, CA 91125, USA \\
$^3$The Henry A. Rowland Department of Physics and Astronomy, Johns Hopkins University, 3400 North Charles Street, \\ Baltimore, MD 21218, USA \\
$^4$Observational Cosmology Lab, Code 665, NASA Goddard Space Flight Center, Greenbelt, MD 20771, USA}

\date{\today}
             
\pagerange{\pageref{firstpage}--\pageref{lastpage}} \pubyear{2014}

\maketitle

\label{firstpage}

\begin{abstract}
Using 1\,cm and 3\,mm CARMA and 2\,mm GISMO observations, we follow up the first extragalactic detection of anomalous microwave emission (AME) reported by \citet{Murphy+etal_2010} in an extranuclear region (Enuc. 4) of the nearby face-on spiral galaxy NGC\,6946. We find the spectral shape and peak frequency of AME in this region to be consistent with models of spinning dust emission. However, the strength of the emission far exceeds the Galactic AME emissivity given the abundance of polycyclic aromatic hydrocarbons (PAHs) in that region. Using our galaxy-wide 1\,cm map (21\arcsec resolution), we identify a total of eight 21\arcsec x21\arcsec regions in NGC\,6946 that harbour AME at $>95\%$ significance at levels comparable to that observed in Enuc. 4. The remainder of the galaxy has 1\,cm emission consistent with or below the observed Galactic AME emissivity per PAH surface density. We probe relationships between the detected AME and dust surface density, PAH emission, and radiation field, though no environmental property emerges to delineate regions with strong versus weak or non-existent AME. On the basis of these data and other AME observations in the literature, we determine that the AME emissivity per unit dust mass is highly variable. We argue that the spinning dust hypothesis, which predicts the AME power to be approximately proportional to the PAH mass, is therefore incomplete.
\end{abstract}

\begin{keywords}
ISM: dust -- radio continuum: ISM
\end{keywords}

\section{Introduction}
Anomalous Microwave Emission (AME) is dust-correlated emission observed between $\sim20 - 50$\,GHz that cannot be accounted for by extrapolating the thermal dust emission to low frequencies. First detected as an emission excess in the microwave \citep{Kogut+etal_1996, deOliveiraCosta+etal_1997, Leitch+etal_1997}, AME is now thought to arise from electric dipole emission from very small grains with size $\sim3.5 - 10$\,\text{\AA} that spin rapidly due to the action of systematic torques in the interstellar medium \citep{Draine+Lazarian_1998b, Hoang+Draine+Lazarian_2010, Ysard+Verstraete_2010, Silsbee+AliHaimoud+Hirata_2011}. Subsequent Galactic observations have been well-fit with the inclusion of a spinning dust component \citep[e.g.][]{Miville-Deschenes+etal_2008, Planck_AME}.

\citet{Murphy+etal_2010} reported the first extragalactic detection of AME nnear the star-forming extra-nuclear region of the spiral galaxy NGC 6946 (hereafter Enuc. 4). NGC 6946, located at a distance of 6.8 Mpc \citep{Karachentsev+etal_2000}, is known for its intense star formation, having hosted at least nine supernovae in the last century \citep{Prieto+etal_2008} and having a star formation rate (SFR) of 7.1\,$M_\odot$ yr$^{-1}$ \citep{Kennicutt+etal_2011}. Follow-up observations of Enuc. 4 by the Arcminute Microkelvin Imager (AMI) support the presence of an AME component in this source \citep{AMI_6946}. 

To date, additional extragalactic detections have proven elusive. \emph{Planck} observations of the Small Magellanic Cloud (SMC) have shown evidence for an AME component \citep{Planck_SMC, Draine+Hensley_2012}, but the interpretation is complicated by additional excess emission possibly arising from the large grains. Despite microwave observations of other dusty, star-forming galaxies including Andromeda, only upper limits on an AME component have been placed \citep{Peel+etal_2011, Planck_Andromeda}. The difficulty in finding extragalactic AME has hindered our ability to probe variations of AME properties within a galaxy, particularly identifying the hallmarks of regions with strong emission.

In this work, we use high resolution data from the Combined Array for Research in Millimeter-wave Astronomy (CARMA) at 1\,cm and 3\,mm to better constrain the spectral energy distribution (SED) of Enuc. 4. We combine these CARMA observations with data from \emph{Spitzer}, \emph{Herschel}, the Goddard IRAM Superconducting 2 Millimeter Observer (GISMO), the Very Large Array (VLA), and the Westerbork Synthesis Radio Telescope (WSRT) to create composite SEDs that allow us to decompose the emission into components from thermal dust, synchrotron, free-free, and AME. We compare the AME found in Enuc. 4 to that observed in the Galaxy and assess the viability of the spinning dust mechanism for explaining the emission. 

The CARMA 1\,cm map provides coverage of the whole galaxy, enabling a search for AME beyond Enuc. 4. As we find multiple additional regions with AME, we use the high-resolution dust model fitting of NGC\,6946 by \citet{Aniano+etal_2012} to explore correlations between the AME and local properties such as dust surface density, PAH emission, and the intensity of the local radiation field.

We have organised the paper as follows: in Section 2 we describe the data used in this analysis and our data reduction; in Section 3 we perform fits to the SEDs using models for emission from thermal dust, free-free, synchrotron, and AME and analyse the results of the fits for particular regions of the Galaxy; in Section 4 we analyse galaxy-wide correlations between AME and local galaxy properties and comment on the nature of the AME observed in NGC 6946 versus Galactic observations; in Section 5 we discuss the implications of our observations for spinning dust theory; and in Section 6 we summarise our conclusions.

\section{Data Reduction and SED Synthesis}
In this Section we describe the datasets used in our analysis. The properties of all maps are summarised in Table~\ref{table:data}.

\subsection{CARMA Data}
\begin{figure*}
    \centering
      	\scalebox{0.50}{\includegraphics{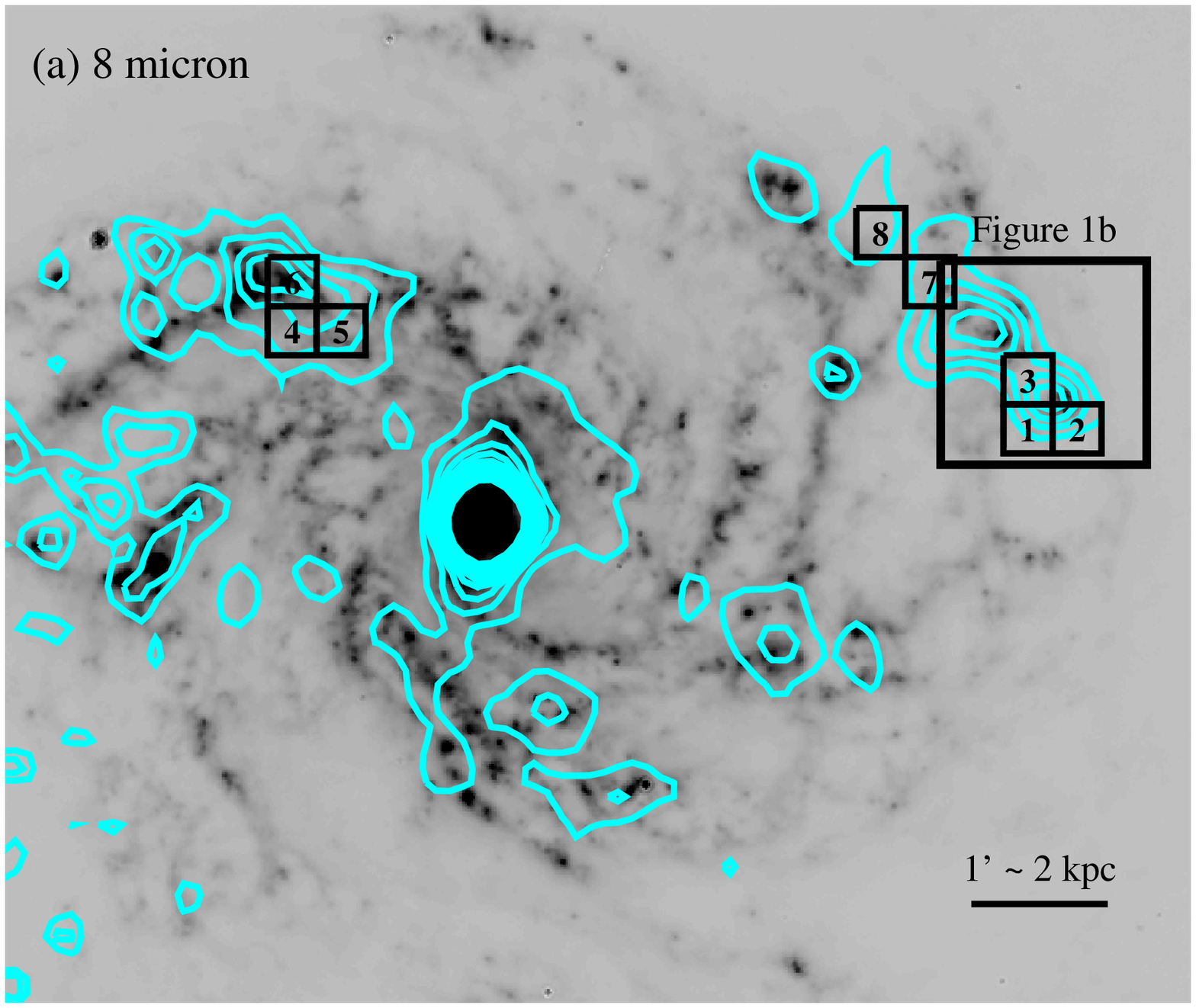}}
        \scalebox{0.40}{\includegraphics{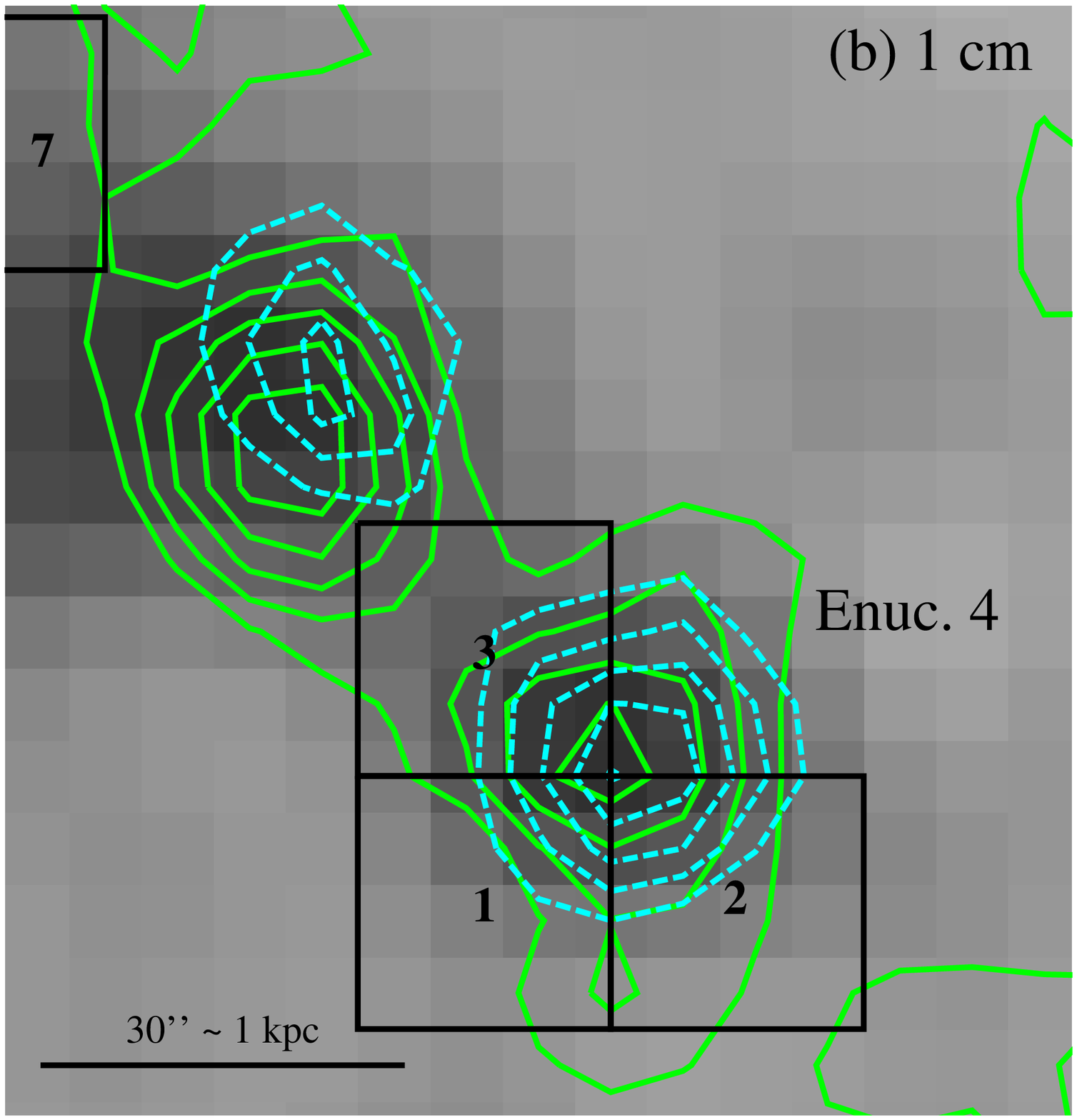}}   
    \caption{\emph{Left}: We overlay the emission contours from the 1\,cm map on the IRAC 8\,$\mu$m image, where the ten contour levels are linearly spaced between the $3\sigma$ noise level of 0.048\,MJy/sr and 0.48\,MJy/sr. The eight pixels with significant AME detections are shown in black boxes with numbering corresponding to Table~\ref{table:ame_detections}. \emph{Right}: We overlay the emission contours from the 2\,mm (green solid) and 3\,mm (cyan dashed) maps on the 1\,cm image in the Enuc. 4 region (indicated by the solid rectangle in the left panel), demonstrating excellent coincidence of the emission in all three maps. The five 2\,mm contours are linearly spaced between 0.05 and 0.2\,MJy/sr while the five 3\,mm contours are linearly spaced between 0.04 and 0.13\,MJy/sr.} \label{fig:carma_images} 
\end{figure*}

CARMA consists of 23 antennas-- six 10\,m, nine 6\,m, and eight 3.5\,m. The CARMA 3\,mm data were taken with 6 and 10\,m antennas in the CARMA E configuration with baseline lengths between 8.5 and 66\,m in July 2010. The map has a resolution of 9.1'' and an rms noise level of 0.017\,MJy/sr. The CARMA 1\,cm data were taken with the eight 3.5\,m Sunyaev-Zel'dovich Array antennas in October 2010 and July 2011 using both the SH and SL configurations. The map has a resolution of 18.8'' and an rms noise level of 0.016\,MJy/sr. The data were reduced with MIRIAD \citep{Sault+Teuben+Wright_2011} using Mars, Jupiter, and Uranus as flux calibrators and assuming a 10\% calibration uncertainty. A primary beam correction was applied to both maps. We present the emission contours from these maps in Figure~\ref{fig:carma_images}.

\subsection{GISMO Data}
GISMO is a Transition Edge Sensor bolometer camera on the IRAM 30\,m telescope that operates in the 2\,mm atmospheric window \citep{Staguhn+etal_2008, Staguhn+etal_2014}. A total of 1.5 hours of integration time was obtained on Enuc 4 in April 2013 at 17\arcsec resolution. The data were reduced using the CRUSH software package ``deep mode'' \citep{Kovacs_2008}, which includes large-scale filtering to mitigate atmospheric effects.

\subsection{Ancillary Data}
\subsubsection{IR Data}
NGC\,6946 was part of both the \emph{Spitzer} Infrared Nearby Galaxies Survey (SINGS) \citep{SINGS} and Key Insights on Nearby Galaxies: a Far-Infrared Survey with \emph{Herschel} (KINGFISH) \citep{Kennicutt+etal_2011} projects, and thus has extensive \emph{Spitzer} and \emph{Herschel} observations. We use the same dataset detailed in \citet{Aniano+etal_2012} consisting of 3.6, 4.5, 5.8, and 8.0\,$\mu$m IRAC images and 24 and 70\,$\mu$m MIPS images from SINGS, and 70, 100, and 160\,$\mu$m PACS images and a 250\,$\mu$m SPIRE image from KINGFISH. We do not use the 160\,$\mu$m MIPS image or the SPIRE 350 or 500\,$\mu$m images due to insufficient angular resolution. All of the remaining images have resolution better than our 18.8\arcsec CARMA 1\,cm map.

\subsubsection{Radio Data}
We employ the same radio data as \citet{Murphy+etal_2010}. These consist of the 1.4\,GHz radio map from the WSRT-SINGS survey \citep{Braun+etal_2007} with a 14\arcsec x12.5\arcsec beam and VLA maps at 1.5, 1.7, 4.9, and 8.5\,GHz, all with 15\arcsec x15\arcsec beams \citep{Beck_2007}.

\begin{table*}
\caption{Data Summary} 
\label{table:data}
  \begin{center}
    \begin{tabular}{cccccc} \hline \hline   
Instrument & $\lambda$ & FWHM & $\sigma_{\rm RMS}$ & $\sigma_{\rm Calibration}$ & Reference \\
 & ($\mu$m) & \arcsec & (MJy/sr) & (\%) & \\ \hline
IRAC & 3.6 & 1.90 & 0.0093 & 5 & \citet{Aniano+etal_2012} \\
IRAC & 4.5 & 1.81 & 0.0093 & 5 & \citet{Aniano+etal_2012} \\
IRAC & 5.8 & 2.11 & 0.0274 & 5 & \citet{Aniano+etal_2012} \\
IRAC & 8.0 & 2.82 & 0.0356 & 5 & \citet{Aniano+etal_2012} \\
MIPS & 24 & 6.43 &  0.0328 & 5 & \citet{Aniano+etal_2012} \\
MIPS & 70 & 18.7 & 0.334 & 10 & \citet{Aniano+etal_2012} \\
PACS & 70 & 5.67 & 2.52 & 10 & \citet{Aniano+etal_2012} \\
PACS & 100 & 7.04 & 2.22 & 10 & \citet{Aniano+etal_2012} \\
PACS & 160 & 11.2 &  1.53 & 20 & \citet{Aniano+etal_2012} \\
SPIRE & 250 & 18.2 & 0.873 & 15 & \citet{Aniano+etal_2012} \\
GISMO & 2000 & 21.0 & 0.060 & 10 & This work \\
CARMA & 3000 & 9.1 & 0.017 & 10 & This work \\
CARMA & 10000 & 18.8 & 0.016 & 10 & This work \\
VLA & 30000 & 14 & 0.010 & 5 & \citet{Braun+etal_2007} \\
VLA & 60000 & 14 & 0.011 & 5 & \citet{Braun+etal_2007} \\
VLA & 18000 & 14 & 0.006 & 5 & \citet{Braun+etal_2007} \\
VLA & 20000 & 14 & 0.006 & 5 & \citet{Braun+etal_2007} \\
WSRT & 22000 & 15 & 0.004 & 5 & \citet{Beck_2007} \\ \hline
    \end{tabular}
  \end{center}
\end{table*}

\subsection{SED Synthesis}
We combined all data by convolving each image with a Gaussian kernel to a common 21\arcsec resolution \citep{Aniano+etal_2011}. This corresponds to a physical scale of about 700\,pc at the distance of NGC\,6946.

\section{Enuc. 4}
The new observations at 1\,cm and 3 and 2\,mm allow us to revisit the SED of Enuc. 4 obtained by \citet{Murphy+etal_2010} and place tighter constraints on the frequency-dependence of the AME. In particular, the inclusion of the 2 and 3\,mm data allow us to better constrain the sum of the AME and thermal dust contributions and verify the presence of a spectral peak in the 30\,GHz range. We extract the flux density in a 21\arcsec diameter circular aperture centred on the Enuc. 4 position given by \citet{Murphy+etal_2010}.

\subsection{Fits Without Spinning Dust}
We first attempt to fit the Eunc. 4 SED without invoking a spinning dust component. We model the synchrotron emission as a power law with unknown normalisation and index, i.e.

\beq
F_\nu^{\rm sync} = A \left(\frac{\nu}{1.36\ {\rm GHz}}\right)^{-\alpha_{\rm sync}}
~~~.
\eeq
We adopt two fitting strategies for the synchrotron slope. Guided by the results of \citet{Niklas+etal_1997}, in the first approach we put a Gaussian prior with mean 0.83 and standard deviation 0.13 on $\alpha_{\rm sync}$. This is a safeguard against fits that yield unphysically low values of $\alpha_{\rm sync}$ to compensate enhanced 1 cm emission. In the second, we simply draw values for $\alpha_{\rm sync}$ uniformly between 0.5 and 1.5. Our results are robust to this choice, and thus we adopt the latter as the default for its simplicity. To fit the amplitude, we expect that the synchrotron comprises the majority of the emission at $22$\,cm and therefore draw the synchrotron normalisation $A$ uniformly between 0 and the $5\sigma$ upper-limit on the $22$ cm data point. 

It is unclear whether a power law is a good description of the synchrotron emission over the frequency range of interest. In particular, the Galactic synchrotron index has been observed to steepen by $\sim0.3$ between 13 and 1\,cm in analyses of WMAP and \emph{Planck} data \citep[e.g.][]{Bennett+etal_2003, Planck_synch_steep}. Such steepening is expected when electrons with requisite energy are no longer able to be produced. Our data have insufficient frequency sampling to meaningfully constrain a break frequency or the resulting index. Therefore, we consider the extreme case in which the synchrotron produces no emission below 3\,cm. Although this will not substantially affect the modeling of the emission in Enuc. 4, steepening is required to explain the SED of other regions of the galaxy, as discussed in detail in Section~\ref{sec:galaxy}.

We model the free-free emission as a power law with fixed index -0.12 and unknown normalisation, i.e.

\beq
F_\nu^{\rm ff} = B \left(\frac{\nu}{30\,{\rm GHz}}\right)^{-0.12}
~~~.
\eeq

Because we possess the full infrared SED of the galaxy, we can constrain the free-free emission based on the empirical relations from \citet{Murphy+etal_2012} connecting the thermal emission, 24\,$\mu$m flux density, and total IR luminosity to the star formation rate. Combining their Equations~6 and~14 and assuming an electron temperature of $10^4$\,K, we obtain

\beq
 \label{eq:ff_24um}
 F_{\nu, \rm 1\,cm}^{\rm ff} = 4.74\times10^{-3} F_{\nu, 24\,\mu{\rm m}}
 ~~~.
 \eeq
Likewise, combining their Equations~6 and~15, we obtain,
 \beq
 \label{eq:ff_Lir}
 \left(\nu L_\nu\right)_{\rm 1\,cm}^{\rm ff} = 1.46\times10^{-6} L_{\rm IR}
 ~~~.
 \eeq
The total IR luminosity from dust has been estimated for each pixel by \citet{Aniano+etal_2012} (their Equations~14 and 21) by fitting the infrared SED with the \citet{Draine+Li_2007} dust model, and we use this value as $L_{\rm IR}$ here.

We adopt two approaches to fit the free-free. In the first, we draw B uniformly between half the minimum and twice the maximum values computed with Equations~\ref{eq:ff_24um} and~\ref{eq:ff_Lir}, which should accommodate both variations in electron temperature and intrinsic scatter in the relations \citep{Murphy+etal_2011, Murphy+etal_2012}. In the second, we draw uniformly between 0 and the 5$\sigma$ upper limit on the 1\,cm emission. We find that our results are robust to choice of prior and adopt the latter method as the default for its simplicity.

Enuc. 4 has a $24\,\mu$m flux density of $5.1\times10^{-2}$\,Jy, implying $2.4\times10^{-4}$\,Jy of free-free emission at 1\,cm. We note immediately that this is far below the total observed 1\,cm flux density of $1.3\times10^{-3}$\,Jy.

For the thermal dust component, we assume a power law of the form

\begin{equation}
F_\nu = C \left(\frac{\nu}{150\,{\rm GHz}}\right)^{\beta}
~~~,
\end{equation}
where we fix $\beta = 3.65$ in accord with measurements of the thermal dust SED by \emph{Planck} and draw $C$ uniformly between 0 and the 5$\sigma$ upper limit on the 2\,mm (150\,GHz) data point.

To perform the fit, we make $2\times10^5$ draws from the prescribed distribution for each parameter, then compute the likelihood $\mathcal{L}$ of each model explicitly:

\begin{equation}
\ \mathcal{L} \propto \prod\limits_i e^{-\frac{\left(\hat{F}_i - F_i\right)^2}{2\sigma_i^2}}
~~~,
\end{equation}
where the $\hat{F}_i$ is the flux density estimated by the model at frequency $i$ and $F_i$ is the data at frequency $i$ that has error $\sigma_i$. The proportionality constant has been omitted since we compare only the relative likelihood of models. We find the 1D confidence intervals for each parameter by marginalising over the others. We obtain nearly identical results on repeating this process, indicating that the number of draws is sufficiently high to sample to the parameter space.

Figure~\ref{fig:enuc4_fit}a gives the fit of these three components to the data, which is poor. Additionally, the required free-free emission at 1\,cm is $7.6\times10^{-4}$ Jy, a factor of three above the estimate based on $F_{\nu, 24\,\mu{\rm m}}$.

\subsection{Spinning Dust Fits}
\label{sec:spdfit}
We perform a second fit employing an additional component from spinning dust emission. Spinning dust emission is a complicated process that depends upon the size, shape, and charge of the emitting grains as well as the environmental conditions such as gas temperature, molecular fraction, ionisation state, and the intensity of the radiation field. Fitting a model that varies all of these parameters is well beyond the capabilities of the data, thus we seek instead a simple prescription for the emission. Following \citet{Draine+Hensley_2012}, we parametrise the spinning dust emission as

\beq
\label{eq:spd_param}
F_{\nu}^{\rm sd} = D \left(\frac{\nu}{\nu_0}\right)^2 \mathrm{exp}\left[1 -
  \left(\frac{\nu}{\nu_0}\right)^2\right]
~~~,
\eeq
where the peak frequency $\nu_0$ and the amplitude $D$ are free parameters. We restrict $\nu_0$ to be between 10 and 70\,GHz and $D$ to be between 0 and the $5\sigma$ upper limit on the 1\,cm data point. Although the model is a simplification of the underlying physics, it is nevertheless instructive to compare both the amplitude and peak frequency obtained from the fits to those of similar studies performed in the Galaxy.

\begin{figure}
    \centering
        \scalebox{0.42}{\includegraphics{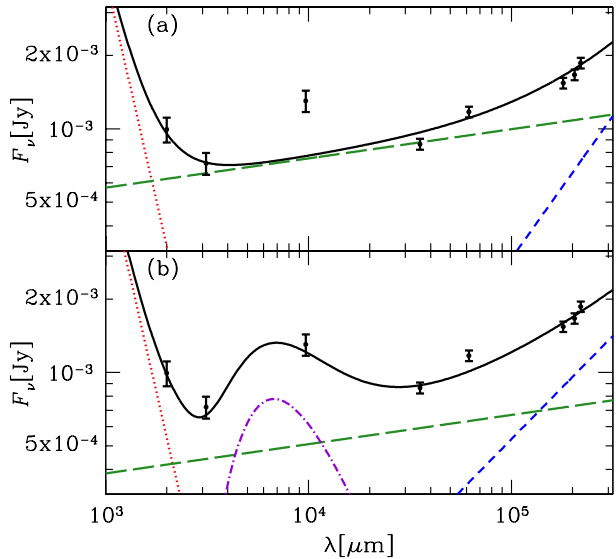}}
    \caption{The fit to the long wavelength Enuc. 4 SED using only free-free (long-dash green), synchrotron (short-dash blue), and thermal dust emission (dotted red) is plotted in panel (a) while the fit in panel (b) includes a contribution from spinning dust emission (dot-dashed violet). The total emission for each model is plotted with a solid black line. The goodness of fit is improved considerably with the inclusion of an AME component with peak frequency 44\,GHz and amplitude 0.76\,mJy at peak.} \label{fig:enuc4_fit} 
\end{figure}

Using the formalism outlined above, we obtain $D = 0.76\pm0.23$\,mJy and $\nu_0 = 44\pm6$\,GHz, corresponding to a 30\,GHz flux density of 0.61\,mJy. The free-free flux density at 1\,cm is $5.1\times10^{-4}$\,Jy, a factor of two above the estimate from $F_{\nu, 24\,\mu{\rm m}}$ but less than what was required by the fit with no spinning dust emission. Figure~\ref{fig:enuc4_fit}b presents the much-improved fit using this model. To quantify the improvement, we perform the likelihood ratio test by constructing the test statistic $\mathcal{D}$:

\begin{equation}
\mathcal{D} = -2 \ln{\mathcal{L}_1} + 2 \ln{\mathcal{L}_2}
~~~,
\end{equation}
where $\mathcal{L}_1$ and $\mathcal{L}_2$ are the likelihoods of the best fit models without and with a spinning dust component, respectively. $\mathcal{D}$ follows a $\chi^2$ distribution with the number of degrees of freedom equal to the difference in the number of free parameters in the two models, here two. We obtain $\mathcal{D} = 13.8$, disfavouring the model with no spinning dust emission with $p \simeq 10^{-3}$. 

\subsection{Thermal Dust Fits}
Having fit the radio data, we now perform a more realistic modeling of the thermal dust emission. We follow the fitting strategy of \citet{Aniano+etal_2012} and employ the \citet{Draine+Li_2007} dust model. This model includes populations of silicate and carbonaceous grains, including a PAH population, heated by a distribution of radiation field intensities such that the differential dust mass d$M_d$ heated by starlight intensities between $U$ and $U$+d$U$ is given by

\begin{equation}
\frac{1}{M_d} \frac{\mathrm{d}M_d}{\mathrm{d}U} = \left(1 - \gamma\right)\delta\left(U - U_{\rm min}\right) + \gamma \frac{\alpha - 1}{U_{\rm min}^{1 - \alpha} + U_{\rm max}^{1 - \alpha}} U^{-\alpha}
\end{equation}
for $U_{\rm min} < U < U_{\rm max}$. The heating spectrum is assumed to be the interstellar radiation field determined by \citet{MMP83} scaled by a constant factor $U$. $U_{\rm min}$, $U_{\rm max}$, $\gamma$, and $\alpha$, are free parameters of the model to be fit. The PAH emission features require the addition of the parameter $q_{\rm PAH}$ defined as the total mass fraction of the dust in PAHs. Following \citet{Aniano+etal_2012}, we set $U_{\rm max} = 10^7$ since most pixels in NGC\,6946 were best fit by this value and the model is in general relatively insensitive to the value of this parameter \citep{Draine+etal_2007}. We define $\bar{U}$ as the dust mass-weighted mean starlight intensity heating the dust.

\begin{figure}
    \centering
        \scalebox{0.4}{\includegraphics{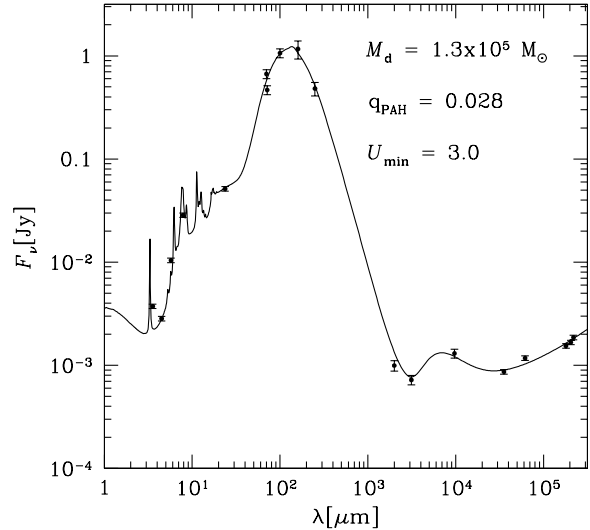}}  
    \caption{The full SED of Enuc. 4 extracted from a 21\arcsec diameter circular aperture (corresponding to a physical scale of $\simeq 700$\,pc) with the best fit model overlaid.} \label{fig:enuc4_dustfit} 
\end{figure}

We fit this model to the Enuc. 4 SED 3\,mm and shortward after subtracting the best fit synchrotron, free-free, and synchrotron as determined in Section~\ref{sec:spdfit}. The thermal dust fit presented in Figure~\ref{fig:enuc4_dustfit} has fit parameters $q_{\rm PAH} = 0.028$, $M_d = 1.3\times10^5\,M_\odot$, $U_{\rm min} = 3$, $\alpha = 1.7$, and $\gamma = 1.1\times10^{-3}$. $f_{\rm PDR}$, the fraction of the dust luminosity radiated from regions with $U > 100$, has a value of 0.18.

\subsection{Comparisons to Spinning Dust Models}
For reasonable assumptions about the size distribution and electric dipole moments of small grains, \citet{Draine+Lazarian_1998a, Draine+Lazarian_1998b} calculated a spinning dust emissivity at 30\,GHz of $1\times10^{-17}$\,Jy sr$^{-1}$ cm$^2$ H$^{-1}$, consistent with the observations of \citet{Kogut+etal_1996}, \citet{deOliveiraCosta+etal_1997}, and \citet{Leitch+etal_1997}, though the scatter was large. Subsequent observations of the H$\alpha$-correlated AME \citep{Dobler+Draine+Finkbeiner_2009}, the Perseus molecular cloud \citep{Tibbs+etal_2010, Tibbs+etal_2011}, an ensemble of Galactic clouds \citep{Planck_clouds}, and the average diffuse ISM \citep{Planck_diffuse2014} have all suggested a lower emissivity of about $3\times10^{-18}$\,Jy sr$^{-1}$ cm$^2$ H$^{-1}$. We adopt this as our benchmark value.

Since spinning dust emission arises from the smallest grains (believed to be PAHs), its flux density should be directly proportional to the mass of small grains present, i.e. $M_{\rm PAH} \equiv q_{\rm PAH} M_d$. It is therefore more appropriate to normalise the AME by PAH mass rather than by hydrogen column. Using the Galactic values of $q_{\rm PAH} = 0.046$ and $M_d/M_H = 0.01$ \citep{Draine+Li_2007}, a spinning dust emissivity of $3\times10^{-18}$\,Jy sr$^{-1}$ cm$^2$ H$^{-1}$, and taking the distance to NGC\,6946 to be 6.8\,Mpc \citep{Karachentsev+etal_2000}, we obtain:

\begin{equation}
\label{eq:ame_theory}
F_\nu^{\rm sd} = 2\times10^{-8}\ \frac{M_{\rm PAH}}{M_\odot} ~\mathrm{Jy}
~~~.
\end{equation}

From the values of $q_{\rm PAH}$ and $M_d$ that were derived from fitting the thermal dust emission, Equation~\ref{eq:ame_theory} implies $F_\nu^{\rm sd} = 0.06$\,mJy of spinning dust emission in Enuc. 4. This is more than a factor of ten below the fit value of $0.61$\,mJy. The emissivity observed in Enuc. 4 corresponds to $2\times10^{-7}$\,Jy/$M_{\rm PAH}$.

Theoretical models of spinning dust emission predict that the peak frequency is sensitive to the environment, in particular the gas density. The peak frequency derived here agrees well with models of the Cold Neutral Medium (CNM) and Warm Ionized Medium (WIM) presented by \citet{Draine+Lazarian_1998a, Draine+Lazarian_1998b}, but disfavor the low frequency ($\simeq 20$\,GHz) peak predicted by models of spinning emission arising from the Warm Neutral Medium (WNM). 


Other large-scale determinations of the AME peak frequency have also yielded results around 40\,GHz. The all-sky AME SED extracted from WMAP \citep{Miville-Deschenes+etal_2008} was fit in part using a CNM component peaking at $\simeq 40$\,GHz by \citet{Hoang+Lazarian+Draine_2011} and \citet{Ysard+Miville-Deschenes+Verstraete_2010}. However, both models also required a dominant WNM component peaking at $\simeq$23\,GHz that we do not see here. Additionally, H$\alpha$-correlated AME was found to peak around 40\,GHz \citep{Dobler+Draine+Finkbeiner_2009}, consistent with spinning dust in the WIM. Stepping outside the Galaxy, the millimeter SMC SED is well-fit using a spinning dust contribution peaking at 40\,GHz \citep{Draine+Hensley_2012}.


We thus conclude that the AME detected in Enuc. 4 has a spectral shape and peak frequency consistent with spinning dust emission from the CNM or WIM as seen in both the Galaxy and the SMC, but is far stronger than Galactic AME per unit PAH mass. It is possible that the dust in NGC\,6946 is substantially different than Galactic dust in composition, leading to an increased spinning dust emissivity. To better understand the evolution of AME strength with environment, we investigate other regions in NGC\,6946 with detected AME in the following Section.

\section{Galaxy-Wide Analysis}
\label{sec:galaxy}
Though centered on Enuc. 4, our CARMA 1\,cm map provides coverage over the entire galaxy. We can therefore leverage the dust fitting by \citet{Aniano+etal_2012} to determine whether correlations exist between the fit AME emissivity and the dust properties of that location. To our knowledge, this is the first time such an analysis has been conducted over such a large scale in an external galaxy.

\subsection{Model Fitting}
We first regridded the maps at each frequency to 21\arcsec x21\arcsec square pixels in order to minimise pixel-to-pixel correlations. The 2 and 3\,mm maps were not used in this analysis as they cover only the region of the galaxy immediately surrounding Enuc. 4. To ensure that we are fitting galaxy and not background pixels, we require a 3$\sigma$ detection in all IR and radio bands \emph{except} the 1\,cm band, where it is useful to place upper limits on the AME contribution. We note that using a $5\sigma$ threshold does not substantially affect the results, but does reduce the number of pixels considered.

\begin{figure*}
    \centering
        \scalebox{1.4}{\includegraphics[width=12.5cm,angle=0,
                 clip=true,trim=1.75cm 2.0cm 0.0cm 7.0cm]{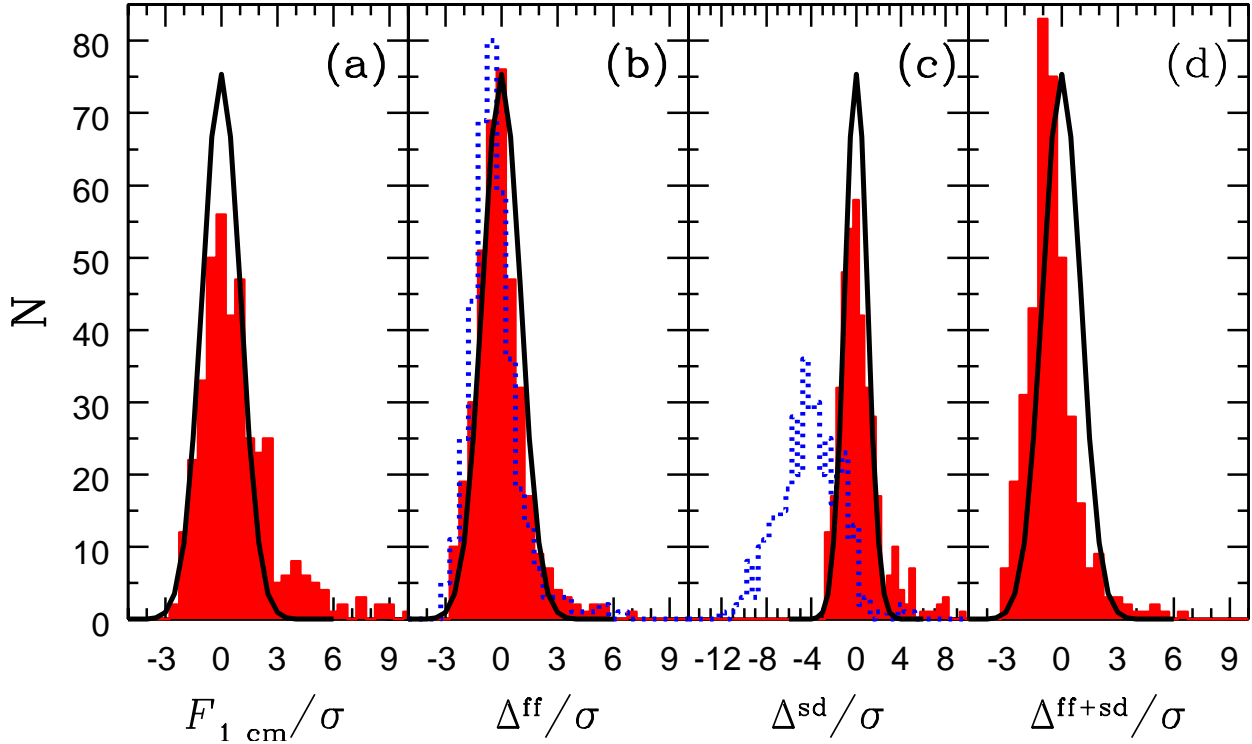}}
    \caption{In all panels, we plot a histogram of the expected number of items in each $0.5\sigma$ bin assuming 382 points are drawn from a Gaussian distribution of mean 0 and standard deviation $\sigma$ (black). In panel (a) we plot the histogram of 1\,cm flux densities normalised by the observational error (red filled). The noise in the map appears adequately estimated. In panel (b), we subtract estimates of the free-free emission from $F_{\nu, \rm 1\,cm}$ and normalise the difference by the 1\,cm error, i.e. $\Delta^{\rm ff} = \left(F_{\nu, \rm 1\,cm} - F^{\rm ff}_{\nu \rm 1\,cm}\right)/\sigma$. The estimate for free-free emission using $F_{\nu 24\,\mu {\rm m}}$ (red filled) appears more consistent with the data than the estimate for free-free emission using $L_{\rm IR}$ (blue dotted). In panel (c), we perform a similar subtraction for estimates of the spinning dust emission from Equation~\ref{eq:ame_theory} (red filled) and the emissivity implied by our fit to Enuc. 4 (blue dotted), i.e. $\Delta^{\rm sd} = \left(F_{\nu, \rm 1\,cm} - F^{\rm sd}_{\nu, \rm 1\,cm}\right)/\sigma$. It is clear that even before accounting for free-free and synchrotron, the emissivity per PAH observed in Enuc. 4 is inconsistent with most regions of the galaxy. Finally in panel (d), we perform a final subtraction using \emph{both} the free-free estimate based on $F_{\nu, 24\,\mu{\rm m}}$ and Equation~\ref{eq:ame_theory}. The fit is generally good, though one or both components may be overestimated. Also, there are clear outliers indicating the entirety of the emission in some pixels has not been accounted for.} \label{fig:hist_1cm} 
\end{figure*}

Before performing a fit to the radio SED, we consider the 1\,cm map alone. In Figure~\ref{fig:hist_1cm}a, we present the histogram of 1\,cm flux densities normalised by the $1\sigma$ error for the 382 pixels meeting the $3\sigma$ criterion. 44 of the pixels have detected 1\,cm emission in excess of $3\sigma$, but most pixels are consistent with the estimated noise level.

Using Equations~\ref{eq:ff_24um} and~\ref{eq:ff_Lir}, we can determine the expected free-free flux density in each of these pixels from both the $24\,\mu$m emission and $L_{\rm IR}$. We plot histograms of the 1\,cm emission with the estimated free-free contribution subtracted in Figure~\ref{fig:hist_1cm}b, where it is clear that our estimate based on $L_{\rm IR}$ systematically overestimates the free-free emission while the estimate based on the 24\,$\mu$m emission is generally consistent with the data. This is not unexpected as the total IR luminosity will also have a contribution from dust heated by an older stellar population not associated with free-free emission around H{\sc ii} regions. The pixels with excess emission after subtraction could have a significant AME contribution or particularly strong free-free or synchrotron emission. SED fitting is required to discriminate between these alternatives.

We estimate the spinning dust emission in each pixel using Equation~\ref{eq:ame_theory}, and for illustration also estimate the emission using a spinning dust emissivity of $2\times10^{-7}$\,Jy/$M_{\rm PAH}$ implied by our fits to Enuc. 4. We present the histogram of AME-subtracted pixels for both emissivities in Figure~\ref{fig:hist_1cm}c. It is clear that the ensemble of pixels are consistent with the spinning dust emission estimated by Equation~\ref{eq:ame_theory}, but the Enuc. 4 emissivity severely overestimates the emission in most pixels even before subtracting a free-free component. In Figure~\ref{fig:hist_1cm}d, we subtract both an estimate of the free-free based on $F_{\nu, 24 \mu{\rm m}}$ (Equation~\ref{eq:ff_24um}) and an estimate of the spinning dust emission based on Equation~\ref{eq:ame_theory}. Although agreement is overall good, the histogram is skewed negative, suggesting either the free-free, AME, or both have been overestimated. Additionally, outliers with residuals in excess of $3\sigma$ remain. This could imply either a strong AME component (similar to what is observed in Enuc. 4), regions in which our free-free estimate breaks down, or regions of particularly strong synchrotron emission.

\begin{figure}
    \centering
    \scalebox{0.4}{\includegraphics{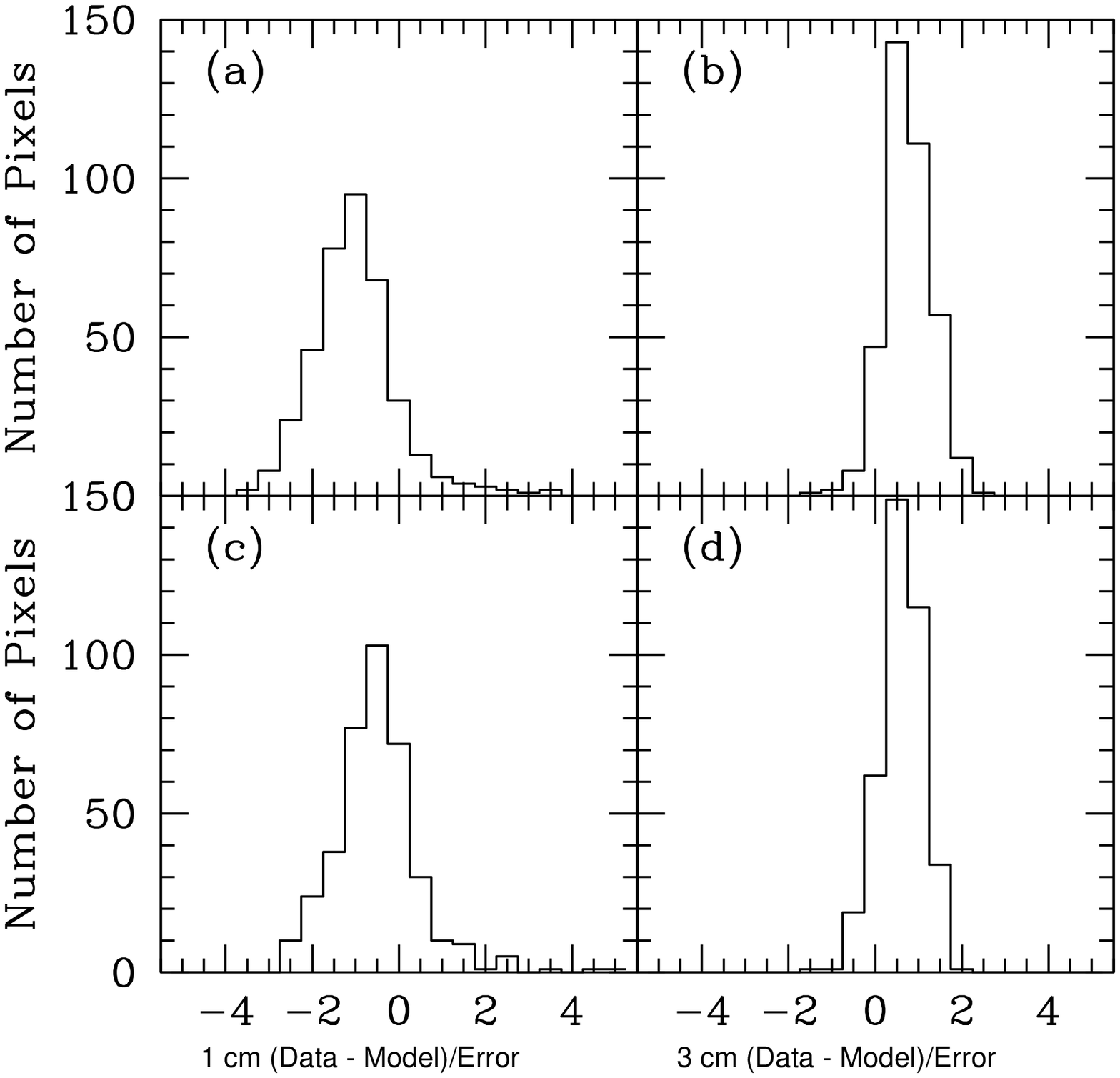}}   
    \caption{Fit residuals at 1 and 3\,cm for radio models without an AME component. Panels (a) and (b) are for a constant synchrotron power law index while panels (c) and (d) are for an index that steepens infinitely at 3\,cm, resulting in no 1\,cm synchrotron emission. Both models strongly indicate the need for an AME component in some regions. The asymmetry in the residuals for the former model suggest that the radio emission is being overestimated at 1\,cm. This asymmetry is somewhat improved in the model with a steepening index, but some negative bias remains.} \label{fig:residuals} 
\end{figure}

To test whether the inclusion of an AME component is warranted by the data, we fit a model with only synchrotron and free-free emission to the SED of each pixel using the likelihood method described above. The 1\,cm residuals from this fit are presented in Figure~\ref{fig:residuals}a, illustrating that the majority of pixels have a residual at 1\,cm consistent with zero, i.e. an AME component is \emph{not} needed to account for the 1\,cm emission in these regions. We note, however, that the histogram of residuals is skewed negative, suggesting that the radio model is overestimating the 1\,cm emission even before the inclusion of an AME component. 

This tension can be alleviated somewhat if the synchrotron power law steepens between 3 and 1\,cm. In Figures~\ref{fig:residuals}c and~\ref{fig:residuals}d, we consider the most extreme case in which the synchrotron component drops to zero below 3\,cm. That the fits continue to overestimate the 1\,cm emission even in this case either suggests that the emission components are not being properly modeled at over wavelengths (particularly 3 and 6\,cm) or that there is a calibration offset between the 1\,cm data and some or all of the remaining points on the SED.

Given the uncertainty in the radio model, we cannot quantify precisely the AME in each pixel. However, as illustrated by Figure~\ref{fig:hist_1cm}c, an emissivity per PAH mass consistent with that observed in Enuc. 4 is already ruled out from the 1\,cm map alone. We can also constrain the AME in pixels with a subdominant radio component where the estimates of AME contribution are relatively insensitive to the synchrotron model, and we focus on these regions in the remainder of this Section.

While the residuals plotted in Figure~\ref{fig:residuals} are helpful for identifying regions in which the free-free and synchrotron models are unable to produce enough emission at 1\,cm to agree with observations, a residual fit will underestimate any true AME at 1\,cm since other parts of the fit will be strained to accommodate a high 1\,cm point. Thus, we perform two additional fits. First, we fit explicitly allowing a contribution from AME at 1\,cm that may be either positive or negative. Second, we fit the data 3\,cm and longward with a radio model and estimate the AME as the fit residual at 1\,cm. The agreement between these two methods is excellent.

Using a non-breaking synchrotron model, we identify three pixels a having AME inconsistent with zero at $3\sigma$ and five additional pixels with AME significant at $2\sigma$. If we allow no synchrotron contribution at 1\,cm, we find two additional pixels significant at $2\sigma$. The detections and their coordinates are listed in Table~\ref{table:ame_detections}. We note that the pixels significant at $3\sigma$ are in the Enuc. 4 area. Three of the five pixels significant at $2\sigma$ are in the immediate vicinity of another star-forming region studied by \citet{Murphy+etal_2010} (i.e. Enuc. 6). The remaining two are located in the extended region of relatively bright 1\,cm emission to the northeast of Enuc. 4 (see Figure~\ref{fig:carma_images}).

\begin{table*}
\caption{AME Detections} 
\label{table:ame_detections}
  \begin{center}
    \begin{tabular}{ccccccccccccccc} \hline \hline  
Pixel & R.A. & Dec. &
$F_{\nu, \rm 1\,cm}^{\rm AME}$ &  &  &
 & $M_d$ & $q_{\rm PAH}$ &
$L_d$ & $\bar{U}$ & $f_{\rm PDR}$ &
$\alpha$ & $\gamma$ & $\alpha_{\rm sync}$ \\
 & (J2000) & (J2000) &
($\mu$Jy) & $\frac{F_{\nu, \rm 1\,cm}^{\rm AME}}{F_{\nu, \rm 1\,cm}}$
 & $\frac{F_{\nu, \rm 100\,\mu m}}{F_{\nu, \rm 1\,cm}^{\rm AME}}$ &
$\frac{F_{\nu, \rm 8\,\mu m}}{F_{\nu, \rm 24\,\mu m}}$ &
$(10^5\,M_\odot)$ &  & $(10^7 L_\odot)$  &  &  &  &  & \\ \hline
1 &      308.5885 &       60.1654 & $  635\pm 170$ &          0.62 & 1300 &          0.67 &          1.12 &          0.03 &          6.29 &          3.35 &          0.13 &          1.62 &          0.00 &          0.67\\
2 &      308.5768 &       60.1653 & $  443\pm 150$ &          0.61 & 1800 &          0.62 &          0.90 &          0.03 &          6.14 &          3.94 &          0.12 &          1.81 &          0.06 &          0.79\\
3 &      308.5885 &       60.1712 & $  714\pm 200$ &          0.55 & 1600 &          0.48 &          1.31 &          0.03 &          9.30 &          4.22 &          0.17 &          1.78 &          0.01 &          0.56\\
4 &      308.7644 &       60.1771 & $  833\pm 350$ &          0.56 & 2200 &          0.88 &          2.19 &          0.04 &         13.31 &          3.64 &          0.12 &          1.52 &          0.00 &          1.12\\
5 &      308.7527 &       60.1771 & $  730\pm 300$ &          0.57 & 2200 &          0.94 &          2.12 &          0.04 &         11.92 &          3.45 &          0.11 &          1.51 &          0.00 &          1.17\\
6 &      308.7644 &       60.1829 & $  813\pm 360$ &          0.49 & 3500 &          0.74 &          2.59 &          0.04 &         21.49 &          4.90 &          0.14 &          1.58 &          0.00 &          0.93\\
7 &      308.6119 &       60.1829 & $  374\pm 170$ &          0.46 & 1900 &          0.87 &          1.21 &          0.03 &          5.70 &          2.84 &          0.10 &          1.47 &          0.00 &          1.00\\
8 &      308.6236 &       60.1887 & $  364\pm 160$ &          0.63 & 1100 &          1.00 &          1.03 &          0.04 &          3.22 &          1.88 &          0.08 &          1.64 &          0.01 &          1.15 \\ \hline
    \end{tabular}
  \end{center}
\flushleft Note-- RA and Dec are reported at the centre of the 21\arcsec x21\arcsec pixel. Thus, pixels in the same row or column of the gridded image will share an RA or Dec.
\end{table*}

\subsection{AME per $M_{\rm PAH}$}
\begin{figure}
    \centering
    \scalebox{0.4}{\includegraphics{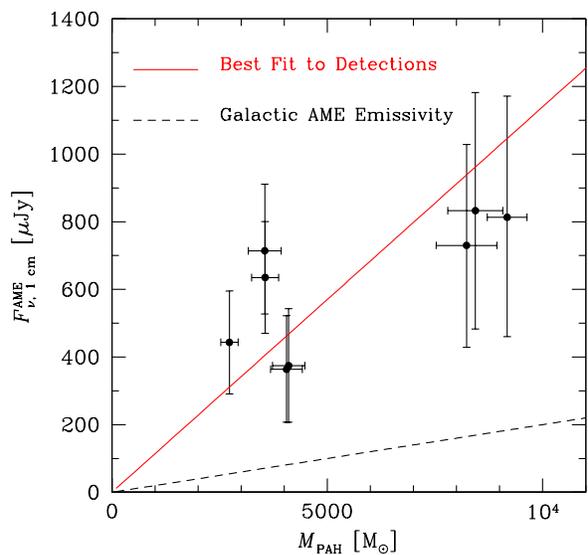}}   
    \caption{The PAH mass in each pixel in Table~\ref{table:ame_detections} is plotted against the fit AME flux density in that pixel. If the AME is arising from rotational emission from spinning grains, the two quantities should be roughly linearly correlated. We overplot the theoretical curve assuming linear correlation and using the canonical spinning dust emissivity (black dashed) as well as our best fit emissivity for the ensemble (red solid).} \label{fig:mpah} 
\end{figure}

Using Equation~\ref{eq:ame_theory} and the $q_{\rm PAH}$ and $M_d$ values in each pixel as determined by \citet{Aniano+etal_2012}, we predict the expected AME in each pixel from spinning dust emission. In Figure~\ref{fig:mpah}, we plot the value of $M_{\rm PAH} = q_{\rm PAH} M_d$ in each pixel against the 1\,cm fit AME flux density $F_{\nu, 1\,{\rm cm}}^{\rm AME}$ for the 8 AME detections. The regions are best fit by an emissivity of $\left(1.1\pm0.2\right)\times10^{-7}$\,Jy/$M_{\rm PAH}$, in sharp tension with the $2\times10^{-8}$\,Jy/$M_{\rm PAH}$ typical of Galactic AME. Likewise, this value is too high to be consistent with the AME non-detections in the remainder of the galaxy as illustrated in Figure~\ref{fig:hist_1cm}c. Therefore, if the AME is rotational emission from small spinning grains, the emissivity must vary significantly even at a fixed number of small grains.

\subsection{AME Detections vs. Non-Detections}
\begin{figure*}
    \centering
        \scalebox{1.4}{\includegraphics[width=12.5cm,angle=0,
                 clip=true,trim=0.75cm 1.0cm 0.0cm 10.0cm]{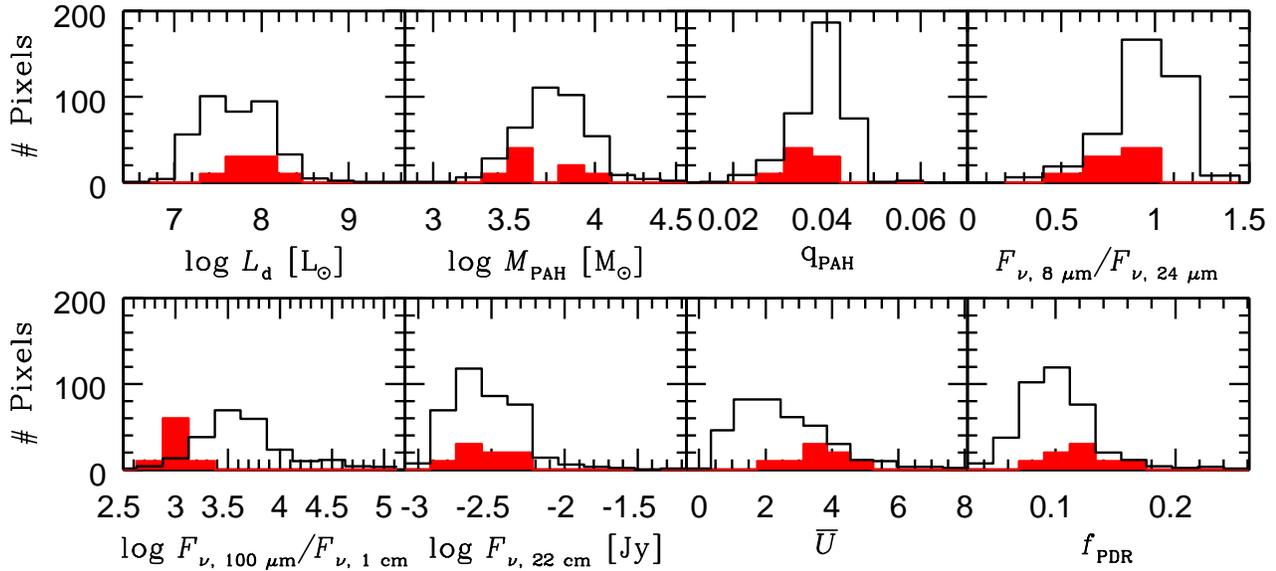}}
    \caption{Histogram of various dust and environmental properties for the AME detections (red filled) and the ensemble of the pixels (black). For readability, we scale the number AME detections up by a factor of ten. No strong trends emerge from contrasting the two populations, although the regions with AME detections appear to favor larger $\bar{U}$ values.} \label{fig:hists} 
\end{figure*}

If some regions of the galaxy produce significantly more AME per PAH mass than others, what sets these regions apart? We contrast the AME detections against the ensemble of pixels in both dust properties ($L_d$, $M_{\rm PAH}$, $q_{\rm PAH}$, $F_{\nu, 8\,\mu{\rm m}}/F_{\nu, 24\,\mu{\rm m}}$) and environmental properties ($F_{\nu, \rm 22 cm}$, $\bar{U}$, $f_{\rm PDR}$) in Figure~\ref{fig:hists}, where for clarity we have scaled up the number of AME detections in each bin by a factor of ten. Of particular note is that these regions appear to have \emph{smaller} $q_{\rm PAH}$ and 8 to 24\,$\mu$m ratio relative to the average value in the galaxy.

Performing a similar investigation within the Perseus molecular cloud, \citet{Tibbs+etal_2011} found that regions with AME tended to have stronger radiation fields than those without. Similarly, the strength of the radiation field was found to be correlated with the AME in the H{\sc ii} region RCW175 by \citet{Tibbs+etal_2012}. We also find a preference for larger $\bar{U}$ and $f_{\rm PDR}$ values among our detections. \citet{Casassus+etal_2006} observed spinning dust emission in the PDR of LDN 1622, suggesting perhaps that spinning dust emission could be enhanced in these regions. We note that the physical scales in our study ($\simeq 1$\,kpc) are much larger than in these studies of Galactic clouds ($\simeq 1 - 10$\,pc). However, if the AME power is robust to environment as predicted by the spinning dust theory, we can make meaningful comparisons regardless of the size of the region. This is particularly true when comparing the AME power per $M_{\rm PAH}$.

The ratio of the AME flux density at 30\,GHz to the 100\,$\mu$m flux density has been a common diagnostic in the literature. \citet{Davies+etal_2006} at intermediate latitudes and \citet{Alves+etal_2010} in the Galactic plane both derive a value of $3\times10^{-4}$; \citet{Todorovic+etal_2010} find a mean ratio of $\left(1.08\pm0.02\right)\times10^{-4}$ in Galactic H{\sc ii} regions; and \citet{Planck_clouds} derive values for Galactic clouds of $\left(2.5\pm0.2\right)\times10^{-4}$ and $\left(5.8\pm0.7\right)\times10^{-4}$ using an unweighted and weighted mean, respectively. An unweighted mean of the values in our Table~\ref{table:ame_detections} yields a ratio of $\left(5.1\pm1.8\right)\times10^{-4}$, consistent with previous results though the variance is large. A large scatter in this quantity is unsurprising given the dependence of the 100\,$\mu$m flux on grain temperature \citep{Tibbs+Paladini+Dickinson_2012}.


We also note that the majority of pixels with a ratio of 100\,$\mu$m to 1\,cm flux density of less than $\sim 1000$ are AME detections. While this in itself is unsurprising since regions with AME will be brighter at 1 cm than regions without, it suggests that this ratio could be a predictor of AME-positive regions using only these two bands. 

There is also a distinct trend for the AME detections to be located in the spiral arms of the galaxy rather than the nuclear region. Additionally, it appears that there are large regions of AME emission that span multiple pixels, suggesting a ``diffuse'' origin of the AME rather than, e.g., compact H{\sc ii} regions.

\subsection{Correlations with AME Intensity}
\begin{figure}
    \centering
        \scalebox{0.7}{\includegraphics[width=12.5cm,angle=0,
                 clip=true,trim=0.5cm 1.0cm 0.0cm 7.0cm]{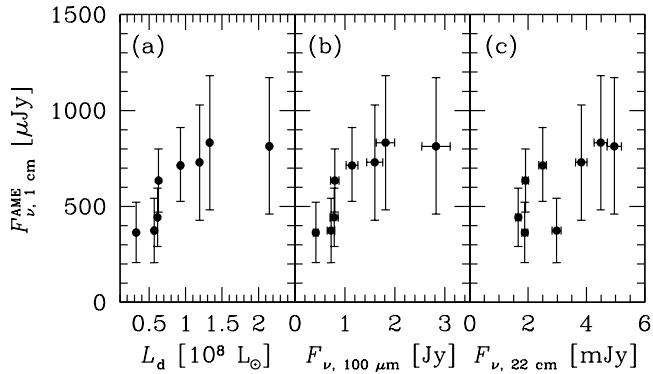}}
    \caption{The fit AME flux density correlates positively with (a) the total dust luminosity $L_d$, (b) the 100\,$\mu$m flux density (and the flux density in all IR bands), and (c) the 22\,$\mu$m flux density (and the flux density in the other radio bands).} \label{fig:ame_correlations} 
\end{figure}

We next examine correlations between the AME intensity and dust and environmental parameters. We find that, as expected, quantities that correlate with dust mass ($M_{\rm PAH}$, $L_d$, $F_{\nu, 100\,\mu{\rm m}}$) correlate with the fit AME intensity (Figures~\ref{fig:mpah}, and~\ref{fig:ame_correlations}). Similarly, we find good correlations with the emission in all of the dust bands, in accord with observations of Galactic clouds by \emph{Planck} \citep{Planck_clouds}. Finally, we also observe positive correlation between the AME intensity and emission in the radio bands, which is expected since both should correlate positively with the gas column. These correlations are summarised in Figure~\ref{fig:ame_correlations}.

To probe variations in AME strength beyond what is expected simply from the amount of gas and dust in that region, we normalise the fit AME flux density by the dust mass in that pixel. This is the normalization suggested by \citet{Tibbs+Paladini+Dickinson_2012} who caution that the 100\,$\mu$m flux, often used instead, is temperature-sensitive. This normalisation is also directly comparable to the results of \citet{Planck_clouds} where the optical depth at 250\,$\mu$m $\tau_{250}$ was used as a normalisation.

\begin{figure*}
    \centering
        \scalebox{1.4}{\includegraphics[width=12.5cm,angle=0,
                 clip=true,trim=1.5cm 1.0cm 0.5cm 10.0cm]{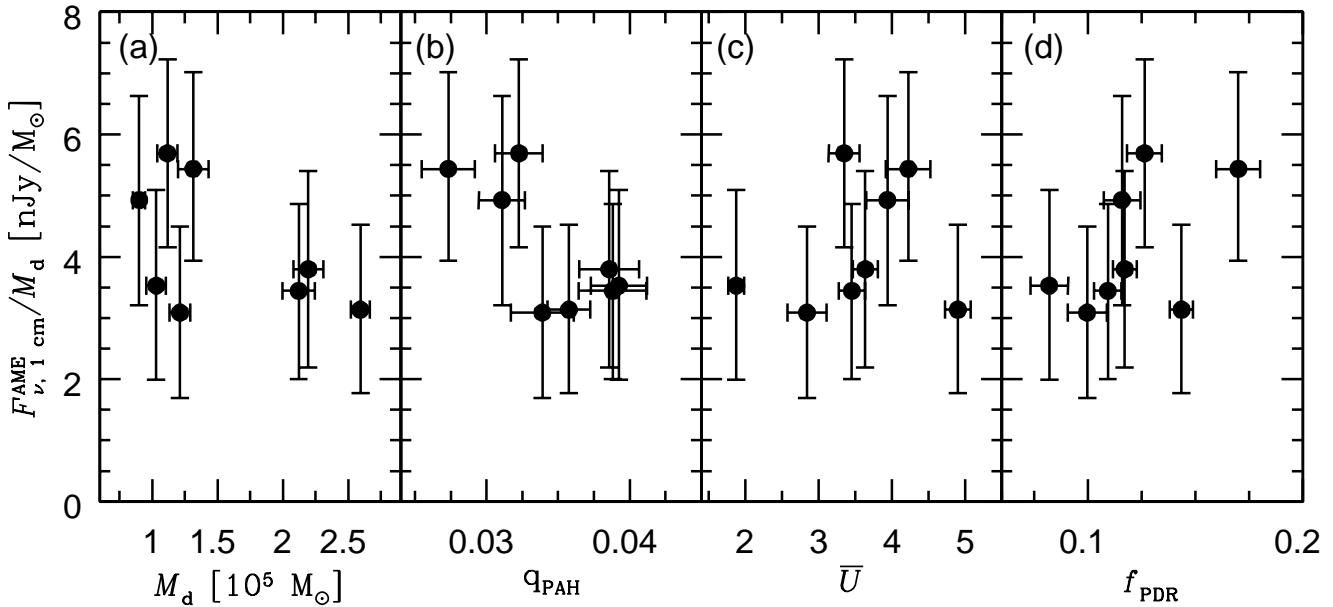}}
    \caption{We examine the correlations between the environment and the strength of the AME per dust mass in the eight regions listed in Table~\ref{table:ame_detections}. As illustrated in panel (a), the AME flux density normalised by the dust mass is consistent with a negative correlation with dust mass, in accord with previous results that AME per column density decreases with column density. In panel (b), we find that a positive linear correlation with $q_{\rm PAH}$ is unfavored by the data. We note that replacing $q_{\rm PAH}$ with the ratio of 8 to 24\,$\mu$m emission produces a nearly identical plot. In the remaining panels, we find no correlations between the normalised AME flux density and (c) $\bar{U}$ or (d) $f_{\rm PDR}$. However, the small sample size precludes drawing strong conclusions from the lack of any correlations.} \label{fig:md_cor} 
\end{figure*}

In Figure~\ref{fig:md_cor}, we plot $M_d$, $q_{\rm PAH}$, $\bar{U}$, and $f_{\rm PDR}$ against the AME flux density normalised by the dust mass. Analysis of WMAP foregrounds by \citet{Lagache_2003} revealed that the AME per column decreases with increasing column, lending credence to the argument that AME is not arising from large grains. \citet{Planck_clouds} reports a sublinear correlation between the AME intensity and the gas column with a power law index of $0.33 \pm 0.04$. Likewise, studying AME in translucent clouds, \citet{Vidal+etal_2011} find that the AME intensity per gas column declines in proportion to the gas column. Among the AME detections in NGC 6946, we find a best fit power law index of $0.59 \pm 0.37$, in accord with the results from \citet{Vidal+etal_2011} and \citet{Planck_clouds}, but limited by our small number of detections.

Although we find tentative agreement with the previous result that the AME is imperfectly correlated with the large grains, we find no better correlation with tracers of small grains. The relationship between $F_{\nu, \rm 1 cm}^{\rm AME}/M_d$ and $q_{\rm PAH}$ is best fit by a power law index of $-0.56\pm0.62$ whereas the value assumed by Equation~\ref{eq:ame_theory} is 1. A similar result is obtained using the 8 to 24\,$\mu$m ratio. 

\citet{Ysard+Miville-Deschenes+Verstraete_2010} find that over the whole sky the AME correlates better with the 12\,$\mu$m emission than the 100\,$\mu$m emission, although it was well-correlated with both. Similarly \citet{Casassus+etal_2006} found the AME in the dark cloud LDN 1622 to be better correlated with maps of the 12 and 25\,$\mu$m emission than the 100\,$\mu$m emission. These trends are not observed in a study of Galactic clouds by \citet{Planck_clouds}, who find that the 12\,$\mu$m map is less correlated with AME than maps at 25, 60, or 100\,$\mu$m. As pointed out by \citet{Tibbs+Paladini+Dickinson_2012}, correlations against 100\,$\mu$m are complicated by the dependence of the 100\,$\mu$m flux on grain temperature. Likewise, the 12\,$\mu$m emission depends not just upon the small grain population but also upon the strength of the radiation field. Taken together with our results, these data do not present a strong case for the association of AME with the smallest grains.

Previous work, both theoretical and observational, suggests an insensitivity of AME amplitude to the radiation field strength. In particular, \citet{Ysard+Verstraete_2010} find no correlation over three orders of magnitude in field strength. We likewise find no correlation, though our AME detections tend to be in regions of above average radiation field intensity. We similarly find no significant correltation of the AME intensity with $f_{\rm PDR}$, though in both cases we are severely limited by the small number of detections.

\section{Discussion}
The AME we observe in Enuc. 4 and several other regions of NGC 6946 is inconsistent with Galactic AME per dust mass even though the AME per 100\,$\mu$m flux density in these regions falls within what has been observed in the Galaxy. Additionally, the emissivity per $M_{\rm PAH}$ in these regions is incompatible with upper limits on the AME in other regions of NGC 6946. Therefore, the data suggest strong variation of the AME intensity at fixed $M_{\rm PAH}$, a result unanticipated by theoretical models of spinning dust emission.

Although they did not estimate the population of small grains explicitly, \citet{Planck_clouds} found a scatter of roughly an order of magnitude in the AME intensity normalised by the optical depth at 250\,$\mu$m in a sample of Galactic clouds. Correlation analysis with a suite of environmental parameters yielded no compelling drivers of increased emissivity, much in accord with our results here.

Such variation is also evident in the Andromeda Galaxy, which has yielded only upper limits on an AME component despite observations with excellent frequency coverage from both WMAP and \emph{Planck} and an ample dust reservoir in the galaxy \citep{Planck_Andromeda}. Recent determinations of the dust content of Andromeda by \citet{Draine+etal_2014} allow us to make a theoretical estimate of the expected spinning dust emission. Inserting the total derived dust mass of $5.4\times10^7$\,M$_\odot$ and the global average value of $q_{\rm PAH} = 0.036$ from \citet{Draine+etal_2014}, and a distance of 744\,kpc \citep{Vilardell+etal_2010} into Equation~\ref{eq:ame_theory} yields an AME flux density of 2.86\,Jy. This value is comparable to the \emph{total} observed 30\,GHz emission of $2.55 \pm 0.44$\,Jy and exceeds the $3\sigma$ upper limit placed on an AME contribution at that frequency of 2.2\,Jy \citep{Planck_Andromeda}. However, the ratio of 100\,$\mu$m to 30\,GHz flux density is $7.1\times10^{-4}$, comparable to our detections and values obtained by \citet{Planck_clouds}.

\citet{Peel+etal_2011} investigate the SEDs of three dusty star-forming galaxies (M82, NGC\,253, and NGC\,4945) for evidence of AME. Using a benchmark AME to 100\,$\mu$m flux density of $3\times10^{-4}$, they find AME flux densities estimated from this ratio far in excess of the observed emission for all three galaxies. This result would be exacerbated using instead the average value of $5\times10^{-4}$ from our Table~\ref{table:ame_detections}. Additionally, due to their excellent frequency coverage between 23 and 143\,GHz, an AME component peaking at a frequency other than 30\,GHz is unlikely to explain the missing flux.

In the context of spinning dust theory, the simplest explanation for variations in emission strength is variation in the abundance of small grains. However, this explanation is inconsistent with our data. We employ $q_{\rm PAH}$  directly from fits to the thermal dust emission to minimise spurious correlation induced by, e.g., varying grain temperatures. The AME is not found preferentially in regions with large values of $q_{\rm PAH}$, the 8 to 24\,$\mu$m ratio, or even $M_{\rm PAH}$. Additionally, a linear scaling of the AME intensity per $M_d$ with $q_{\rm PAH}$ is disfavored by our data at greater than $2\sigma$. If the spinning dust paradigm is correct, and if certain regions of NGC 6946 have AME much stronger than what has been observed in, e.g., Galactic clouds, then an explanation must be furnished for why the small grains in these regions are producing electric dipole radiation in excess of the expected level. Likewise, if the spinning dust emissivity in Andromeda is indeed much weaker than in the Milky Way despite comparable populations of small grains, this must too be explained. 

We examine a few possibilities that may reconcile the variation in emissivity with the theoretical expectations, but none appear able to adequately resolve the tension. First, we have considered only the emission at 30\,GHz, but AME has been observed over a range of peak frequencies from about 20 to 50\,GHz \citep[e.g.][]{Planck_clouds}. It may be possible that regions like Enuc. 4 simply have AME peaking closer to 30\,GHz than the typical diffuse ISM or Galactic clouds. However, if the AME were peaking at 20\,GHz, the 30\,GHz emission would be about two thirds the peak value assuming the spectrum of Equation~\ref{eq:spd_param}. Likewise, if the AME peaked at 40\,GHz, the 30\,GHz amplitude would be lower than the peak amplitude by just over ten per cent. This effect is insufficient to account for the entirety of the variation in NGC\,6946 and not permitted by the multi-frequency observations of Andromeda \citep{Planck_Andromeda} or the star-forming galaxies of \citet{Peel+etal_2011}.

Second, the derived value of $q_{\rm PAH}$ does not correspond precisely to the total PAH fraction by mass. Rather, it is determined from the IR emission features and is therefore weighted toward regions with the high $U$ value. Thus, if there were a reservoir of cold dust that was significantly depleted in PAHs (due, for instance, to grain coagulation), then the derived $q_{\rm PAH}$ would not reflect the true PAH mass of that region. However, the variations in $q_{\rm PAH}$ with environment required to reconcile the data with the theoretical emissivity would appear to be far greater than the observed pixel-to-pixel variations of $q_{\rm PAH}$ itself within the galaxy.

Third, $q_{\rm PAH}$ traces only the abundance of ultrasmall carbonaceous grains. If the AME arises in part or totally from ultrasmall grains of a different composition, such as ultrasmall silicates, then the abundance of AME-producing grains would not be accurately traced by $q_{\rm PAH}$. However it is unclear why these grain populations would not vary cospatially nor does it seem likely that this could account for the entirety of the observed variations.

Finally, there exist mechanisms within the spinning dust theory that can alter the spinning dust emissivity. The typical electric dipole moment of a grain may vary from region to region due, for instance, to variations in grain asymmetry. However, we would expect these variations to be accompanied by environmental variations that explain the differences, such as a more intense radiation field, which we have not seen. 

Taken in conjunction with the previous literature, our study suggests that the AME emissivity per $M_{\rm PAH}$ varies by more than a factor of 10 between the Andromeda Galaxy, the diffuse ISM of the Milky Way, Galactic clouds, and certain extra-nuclear regions of NGC\,6946. Without a clear means of discriminating between regions of particularly high or low AME, the spinning dust hypothesis is at best incomplete.

While we demonstrate the power of coupling AME observations with detailed modelling of the thermal dust component, we are limited by our frequency coverage and sensitivity. Follow-up observations that could better pin down the free-free component would be invaluable for more detailed SED modeling. In addition to mitigating a key uncertainty, such data would also allow exploration of the variations of the AME peak frequency with environmental parameters. Deeper 1\,cm observations of NGC 6946 could better test whether the diffuse ISM of the galaxy has an AME emissivity consistent with Equation~\ref{eq:ame_theory}.

\section{Conclusions}
We have combined new 1\,cm and 3\,mm CARMA observations and 2\,mm GISMO observations with existing IR and radio data to constrain the SED of the AME in NGC\,6946. Our principal conclusions are as follows:

\begin{enumerate}
\item We confirm the detection of AME by \citet{Murphy+etal_2010} in Enuc. 4 and find that the emission is well-fit by a spinning dust component having amplitude 0.9\,mJy and peak frequency 42\,GHz. The peak frequency corresponds well with models of the CNM and WIM. However, the emissivity exceeds the theoretical value by about a factor of ten given its PAH surface density.
\item We find eight regions with AME significant at $> 95\%$ confidence. Multi-frequency follow-up observations of these regions are needed to confirm whether this excess is attributable to spinning dust emission. Such observations could further explore the spatial variations of the spinning dust emission within the galaxy, particularly dependence of the peak frequency on environment.
\item The strength of the AME in these regions is well-correlated with all tracers of dust and gass mass, and we derive a spinning dust emissivity of $1\times10^{-7}$\,Jy/$M_{\rm PAH}$ in these regions, a factor of five higher than the Galactic value.
\item Our results are consistent with previous studies that suggest the AME per unit column density declines with increasing column density. We find no indications that AME correlates with the presence of small grains or the intensity of the radiation field, though these conclusions are limited by a small sample size.
\item The majority of locations in the galaxy are inconsistent with a spinning dust emissivity as strong as that observed in Enuc. 4 given their PAH surface density. This suggests that, if indeed the AME is emission from spinning ultra-small grains, other environmental factors must influence the strength of the emission. However, we find no compelling environmental discriminator between regions with and without detected AME.
\end{enumerate}

\section*{Acknowledgements}
We thank John Carpenter for invaluable assistance reducing the CARMA data and Bruce Draine for extensive feedback that greatly improved the manuscript. We also thank Tim Brandt and Chris White for helpful conversations.

B.H. acknowledges support from the NSF Graduate Research Fellowship under Grant No. DGE-0646086 and NSF grant AST-1408723. The GISMO observations and J.S were supported through NSF ATI grants 1020981 and 1106284.

\bibliography{ngc6946}

\begin{thebibliography}{}
\makeatletter
\relax
\def\mn@urlcharsother{\let\do\@makeother \do\$\do\&\do\#\do\^\do\_\do\%\do\~}
\def\mn@doi{\begingroup\mn@urlcharsother \@ifnextchar[{\mn@doi@}{\mn@doi@[]}}
\def\mn@doi@[#1]#2{\def\@tempa{#1}\ifx\@tempa\@empty
  \href{http://dx.doi.org/#2}{doi:#2}\else \href{http://dx.doi.org/#2}{#1}\fi
  \endgroup}
\def\mn@eprint#1#2{\mn@eprint@#1:#2::\@nil}
\def\mn@eprint@arXiv#1{\href{http://arxiv.org/abs/#1}{{\tt arXiv:#1}}}
\def\mn@eprint@dblp#1{\href{http://dblp.uni-trier.de/rec/bibtex/#1.xml}{dblp:#1}}
\def\mn@eprint@#1:#2:#3:#4\@nil{\def\@tempa {#1}\def\@tempb {#2}\def\@tempc
  {#3}\ifx \@tempc \@empty \let\@tempc\@tempb \let\@tempb\@tempa \fi \ifx
  \@tempb \@empty \def\@tempb{arXiv}\fi \@ifundefined
  {mn@eprint@\@tempb}{\@tempb:\@tempc}{\expandafter \expandafter \csname
  mn@eprint@\@tempb\endcsname \expandafter{\@tempc}}}

\bibitem[\protect\citeauthoryear{{Ade} et~al.,}{{Ade}
  et~al.}{2014}]{Planck_Andromeda}
{Ade} P.~A.~R.,  et~al., 2014, ArXiv e-prints, \href
  {http://adsabs.harvard.edu/abs/2014arXiv1407.5452A} {}

\bibitem[\protect\citeauthoryear{{Alves}, {Davies}, {Dickinson}, {Davis},
  {Auld}, {Calabretta}  \& {Staveley-Smith}}{{Alves}
  et~al.}{2010}]{Alves+etal_2010}
{Alves} M.~I.~R.,  {Davies} R.~D.,  {Dickinson} C.,  {Davis} R.~J.,  {Auld}
  R.~R.,  {Calabretta} M.,   {Staveley-Smith} L.,  2010, \mn@doi [\mnras]
  {10.1111/j.1365-2966.2010.16595.x}, \href
  {http://adsabs.harvard.edu/abs/2010MNRAS.405.1654A} {405, 1654}

\bibitem[\protect\citeauthoryear{{Aniano}, {Draine}, {Gordon}  \&
  {Sandstrom}}{{Aniano} et~al.}{2011}]{Aniano+etal_2011}
{Aniano} G.,  {Draine} B.~T.,  {Gordon} K.~D.,   {Sandstrom} K.,  2011, \mn@doi
  [\pasp] {10.1086/662219}, \href
  {http://adsabs.harvard.edu/abs/2011PASP..123.1218A} {123, 1218}

\bibitem[\protect\citeauthoryear{{Aniano} et~al.,}{{Aniano}
  et~al.}{2012}]{Aniano+etal_2012}
{Aniano} G.,  et~al., 2012, \mn@doi [\apj] {10.1088/0004-637X/756/2/138}, \href
  {http://adsabs.harvard.edu/abs/2012ApJ...756..138A} {756, 138}

\bibitem[\protect\citeauthoryear{{Beck}}{{Beck}}{2007}]{Beck_2007}
{Beck} R.,  2007, \mn@doi [\aap] {10.1051/0004-6361:20066988}, \href
  {http://adsabs.harvard.edu/abs/2007A%26A...470..539B} {470, 539}

\bibitem[\protect\citeauthoryear{{Bennett} et~al.,}{{Bennett}
  et~al.}{2003}]{Bennett+etal_2003}
{Bennett} C.~L.,  et~al., 2003, \mn@doi [\apjs] {10.1086/377252}, \href
  {http://adsabs.harvard.edu/abs/2003ApJS..148...97B} {148, 97}

\bibitem[\protect\citeauthoryear{{Braun}, {Oosterloo}, {Morganti}, {Klein}  \&
  {Beck}}{{Braun} et~al.}{2007}]{Braun+etal_2007}
{Braun} R.,  {Oosterloo} T.~A.,  {Morganti} R.,  {Klein} U.,   {Beck} R.,
  2007, \mn@doi [\aap] {10.1051/0004-6361:20066092}, \href
  {http://adsabs.harvard.edu/abs/2007A%26A...461..455B} {461, 455}

\bibitem[\protect\citeauthoryear{{Casassus}, {Cabrera}, {F{\"o}rster},
  {Pearson}, {Readhead}  \& {Dickinson}}{{Casassus}
  et~al.}{2006}]{Casassus+etal_2006}
{Casassus} S.,  {Cabrera} G.~F.,  {F{\"o}rster} F.,  {Pearson} T.~J.,
  {Readhead} A.~C.~S.,   {Dickinson} C.,  2006, \mn@doi [\apj]
  {10.1086/499517}, \href {http://adsabs.harvard.edu/abs/2006ApJ...639..951C}
  {639, 951}

\bibitem[\protect\citeauthoryear{{Davies}, {Dickinson}, {Banday}, {Jaffe},
  {G{\'o}rski}  \& {Davis}}{{Davies} et~al.}{2006}]{Davies+etal_2006}
{Davies} R.~D.,  {Dickinson} C.,  {Banday} A.~J.,  {Jaffe} T.~R.,  {G{\'o}rski}
  K.~M.,   {Davis} R.~J.,  2006, \mn@doi [\mnras]
  {10.1111/j.1365-2966.2006.10572.x}, \href
  {http://adsabs.harvard.edu/abs/2006MNRAS.370.1125D} {370, 1125}

\bibitem[\protect\citeauthoryear{{Dobler}, {Draine}  \& {Finkbeiner}}{{Dobler}
  et~al.}{2009}]{Dobler+Draine+Finkbeiner_2009}
{Dobler} G.,  {Draine} B.,   {Finkbeiner} D.~P.,  2009, \mn@doi [\apj]
  {10.1088/0004-637X/699/2/1374}, \href
  {http://adsabs.harvard.edu/abs/2009ApJ...699.1374D} {699, 1374}

\bibitem[\protect\citeauthoryear{{Draine} \& {Hensley}}{{Draine} \&
  {Hensley}}{2012}]{Draine+Hensley_2012}
{Draine} B.~T.,  {Hensley} B.,  2012, \mn@doi [\apj]
  {10.1088/0004-637X/757/1/103}, \href
  {http://adsabs.harvard.edu/abs/2012ApJ...757..103D} {757, 103}

\bibitem[\protect\citeauthoryear{{Draine} \& {Lazarian}}{{Draine} \&
  {Lazarian}}{1998a}]{Draine+Lazarian_1998a}
{Draine} B.~T.,  {Lazarian} A.,  1998a, \mn@doi [\apjl] {10.1086/311167}, \href
  {http://adsabs.harvard.edu/abs/1998ApJ...494L..19D} {494, L19}

\bibitem[\protect\citeauthoryear{{Draine} \& {Lazarian}}{{Draine} \&
  {Lazarian}}{1998b}]{Draine+Lazarian_1998b}
{Draine} B.~T.,  {Lazarian} A.,  1998b, \mn@doi [\apj] {10.1086/306387}, \href
  {http://adsabs.harvard.edu/abs/1998ApJ...508..157D} {508, 157}

\bibitem[\protect\citeauthoryear{{Draine} \& {Li}}{{Draine} \&
  {Li}}{2007}]{Draine+Li_2007}
{Draine} B.~T.,  {Li} A.,  2007, \mn@doi [\apj] {10.1086/511055}, \href
  {http://adsabs.harvard.edu/abs/2007ApJ...657..810D} {657, 810}

\bibitem[\protect\citeauthoryear{{Draine} et~al.,}{{Draine}
  et~al.}{2007}]{Draine+etal_2007}
{Draine} B.~T.,  et~al., 2007, \mn@doi [\apj] {10.1086/518306}, \href
  {http://adsabs.harvard.edu/abs/2007ApJ...663..866D} {663, 866}

\bibitem[\protect\citeauthoryear{{Draine} et~al.,}{{Draine}
  et~al.}{2014}]{Draine+etal_2014}
{Draine} B.~T.,  et~al., 2014, \mn@doi [\apj] {10.1088/0004-637X/780/2/172},
  \href {http://adsabs.harvard.edu/abs/2014ApJ...780..172D} {780, 172}

\bibitem[\protect\citeauthoryear{{Hoang}, {Draine}  \& {Lazarian}}{{Hoang}
  et~al.}{2010}]{Hoang+Draine+Lazarian_2010}
{Hoang} T.,  {Draine} B.~T.,   {Lazarian} A.,  2010, \mn@doi [\apj]
  {10.1088/0004-637X/715/2/1462}, \href
  {http://adsabs.harvard.edu/abs/2010ApJ...715.1462H} {715, 1462}

\bibitem[\protect\citeauthoryear{{Hoang}, {Lazarian}  \& {Draine}}{{Hoang}
  et~al.}{2011}]{Hoang+Lazarian+Draine_2011}
{Hoang} T.,  {Lazarian} A.,   {Draine} B.~T.,  2011, \mn@doi [\apj]
  {10.1088/0004-637X/741/2/87}, \href
  {http://adsabs.harvard.edu/abs/2011ApJ...741...87H} {741, 87}

\bibitem[\protect\citeauthoryear{{Karachentsev}, {Sharina}  \&
  {Huchtmeier}}{{Karachentsev} et~al.}{2000}]{Karachentsev+etal_2000}
{Karachentsev} I.~D.,  {Sharina} M.~E.,   {Huchtmeier} W.~K.,  2000, \aap,
  \href {http://adsabs.harvard.edu/abs/2000A%26A...362..544K} {362, 544}

\bibitem[\protect\citeauthoryear{{Kennicutt} Jr. et~al.,}{{Kennicutt}
  et~al.}{2003}]{SINGS}
{Kennicutt} Jr. R.~C.,  et~al., 2003, \mn@doi [\pasp] {10.1086/376941}, \href
  {http://adsabs.harvard.edu/abs/2003PASP..115..928K} {115, 928}

\bibitem[\protect\citeauthoryear{{Kennicutt} et~al.,}{{Kennicutt}
  et~al.}{2011}]{Kennicutt+etal_2011}
{Kennicutt} R.~C.,  et~al., 2011, \mn@doi [\pasp] {10.1086/663818}, \href
  {http://adsabs.harvard.edu/abs/2011PASP..123.1347K} {123, 1347}

\bibitem[\protect\citeauthoryear{{Kogut}, {Banday}, {Bennett}, {Gorski},
  {Hinshaw}  \& {Reach}}{{Kogut} et~al.}{1996}]{Kogut+etal_1996}
{Kogut} A.,  {Banday} A.~J.,  {Bennett} C.~L.,  {Gorski} K.~M.,  {Hinshaw} G.,
   {Reach} W.~T.,  1996, \mn@doi [\apj] {10.1086/176947}, \href
  {http://adsabs.harvard.edu/abs/1996ApJ...460....1K} {460, 1}

\bibitem[\protect\citeauthoryear{{Kov{\'a}cs}}{{Kov{\'a}cs}}{2008}]{Kovacs_2008}
{Kov{\'a}cs} A.,  2008, in Society of Photo-Optical Instrumentation Engineers
  (SPIE) Conference Series. , \mn@eprint {arXiv} {0805.3928},
  \mn@doi{10.1117/12.790276}

\bibitem[\protect\citeauthoryear{{Lagache}}{{Lagache}}{2003}]{Lagache_2003}
{Lagache} G.,  2003, \mn@doi [\aap] {10.1051/0004-6361:20030545}, \href
  {http://adsabs.harvard.edu/abs/2003A%26A...405..813L} {405, 813}

\bibitem[\protect\citeauthoryear{{Leitch}, {Readhead}, {Pearson}  \&
  {Myers}}{{Leitch} et~al.}{1997}]{Leitch+etal_1997}
{Leitch} E.~M.,  {Readhead} A.~C.~S.,  {Pearson} T.~J.,   {Myers} S.~T.,  1997,
  \mn@doi [\apjl] {10.1086/310823}, \href
  {http://adsabs.harvard.edu/abs/1997ApJ...486L..23L} {486, L23}

\bibitem[\protect\citeauthoryear{{Mathis}, {Mezger}  \& {Panagia}}{{Mathis}
  et~al.}{1983}]{MMP83}
{Mathis} J.~S.,  {Mezger} P.~G.,   {Panagia} N.,  1983, \aap, \href
  {http://adsabs.harvard.edu/abs/1983A%26A...128..212M} {128, 212}

\bibitem[\protect\citeauthoryear{{Miville-Desch{\^e}nes}, {Ysard}, {Lavabre},
  {Ponthieu}, {Mac{\'{\i}}as-P{\'e}rez}, {Aumont}  \&
  {Bernard}}{{Miville-Desch{\^e}nes}
  et~al.}{2008}]{Miville-Deschenes+etal_2008}
{Miville-Desch{\^e}nes} M.-A.,  {Ysard} N.,  {Lavabre} A.,  {Ponthieu} N.,
  {Mac{\'{\i}}as-P{\'e}rez} J.~F.,  {Aumont} J.,   {Bernard} J.~P.,  2008,
  \mn@doi [\aap] {10.1051/0004-6361:200809484}, \href
  {http://adsabs.harvard.edu/abs/2008A%26A...490.1093M} {490, 1093}

\bibitem[\protect\citeauthoryear{{Murphy} et~al.,}{{Murphy}
  et~al.}{2010}]{Murphy+etal_2010}
{Murphy} E.~J.,  et~al., 2010, \mn@doi [\apjl] {10.1088/2041-8205/709/2/L108},
  \href {http://adsabs.harvard.edu/abs/2010ApJ...709L.108M} {709, L108}

\bibitem[\protect\citeauthoryear{{Murphy} et~al.,}{{Murphy}
  et~al.}{2011}]{Murphy+etal_2011}
{Murphy} E.~J.,  et~al., 2011, \mn@doi [\apj] {10.1088/0004-637X/737/2/67},
  \href {http://adsabs.harvard.edu/abs/2011ApJ...737...67M} {737, 67}

\bibitem[\protect\citeauthoryear{{Murphy} et~al.,}{{Murphy}
  et~al.}{2012}]{Murphy+etal_2012}
{Murphy} E.~J.,  et~al., 2012, \mn@doi [\apj] {10.1088/0004-637X/761/2/97},
  \href {http://adsabs.harvard.edu/abs/2012ApJ...761...97M} {761, 97}

\bibitem[\protect\citeauthoryear{{Niklas}, {Klein}  \& {Wielebinski}}{{Niklas}
  et~al.}{1997}]{Niklas+etal_1997}
{Niklas} S.,  {Klein} U.,   {Wielebinski} R.,  1997, \aap, \href
  {http://adsabs.harvard.edu/abs/1997A%26A...322...19N} {322, 19}

\bibitem[\protect\citeauthoryear{{Peel}, {Dickinson}, {Davies}, {Clements}  \&
  {Beswick}}{{Peel} et~al.}{2011}]{Peel+etal_2011}
{Peel} M.~W.,  {Dickinson} C.,  {Davies} R.~D.,  {Clements} D.~L.,   {Beswick}
  R.~J.,  2011, \mn@doi [\mnras] {10.1111/j.1745-3933.2011.01108.x}, \href
  {http://adsabs.harvard.edu/abs/2011MNRAS.416L..99P} {416, L99}

\bibitem[\protect\citeauthoryear{{Planck Collaboration} et~al.,}{{Planck
  Collaboration} et~al.}{2011a}]{Planck_SMC}
{Planck Collaboration} et~al., 2011a, \mn@doi [\aap]
  {10.1051/0004-6361/201116473}, \href
  {http://adsabs.harvard.edu/abs/2011A%26A...536A..17P} {536, A17}

\bibitem[\protect\citeauthoryear{{Planck Collaboration} et~al.,}{{Planck
  Collaboration} et~al.}{2011b}]{Planck_AME}
{Planck Collaboration} et~al., 2011b, \mn@doi [\aap]
  {10.1051/0004-6361/201116470}, \href
  {http://adsabs.harvard.edu/abs/2011A%26A...536A..20P} {536, A20}

\bibitem[\protect\citeauthoryear{{Planck Collaboration} et~al.,}{{Planck
  Collaboration} et~al.}{2014a}]{Planck_synch_steep}
{Planck Collaboration} et~al., 2014a, ArXiv e-prints, \href
  {http://adsabs.harvard.edu/abs/2014arXiv1406.5093P} {}

\bibitem[\protect\citeauthoryear{{Planck Collaboration} et~al.,}{{Planck
  Collaboration} et~al.}{2014b}]{Planck_clouds}
{Planck Collaboration} et~al., 2014b, \mn@doi [\aap]
  {10.1051/0004-6361/201322612}, \href
  {http://adsabs.harvard.edu/abs/2014A%26A...565A.103P} {565, A103}

\bibitem[\protect\citeauthoryear{{Planck Collaboration} et~al.,}{{Planck
  Collaboration} et~al.}{2014c}]{Planck_diffuse2014}
{Planck Collaboration} et~al., 2014c, \mn@doi [\aap]
  {10.1051/0004-6361/201323270}, \href
  {http://adsabs.harvard.edu/abs/2014A%26A...566A..55P} {566, A55}

\bibitem[\protect\citeauthoryear{{Prieto} et~al.,}{{Prieto}
  et~al.}{2008}]{Prieto+etal_2008}
{Prieto} J.~L.,  et~al., 2008, \mn@doi [\apjl] {10.1086/589922}, \href
  {http://adsabs.harvard.edu/abs/2008ApJ...681L...9P} {681, L9}

\bibitem[\protect\citeauthoryear{{Sault}, {Teuben}  \& {Wright}}{{Sault}
  et~al.}{2011}]{Sault+Teuben+Wright_2011}
{Sault} R.~J.,  {Teuben} P.~J.,   {Wright} M.~C.~H.,  2011, {MIRIAD:
  Multi-channel Image Reconstruction, Image Analysis, and Display}, \mn@eprint
  {ascl} {1106.007}

\bibitem[\protect\citeauthoryear{{Scaife} et~al.,}{{Scaife}
  et~al.}{2010}]{AMI_6946}
{Scaife} A.~M.~M.,  et~al., 2010, \mn@doi [\mnras]
  {10.1111/j.1745-3933.2010.00878.x}, \href
  {http://adsabs.harvard.edu/abs/2010MNRAS.406L..45S} {406, L45}

\bibitem[\protect\citeauthoryear{{Silsbee}, {Ali-Ha{\"i}moud}  \&
  {Hirata}}{{Silsbee} et~al.}{2011}]{Silsbee+AliHaimoud+Hirata_2011}
{Silsbee} K.,  {Ali-Ha{\"i}moud} Y.,   {Hirata} C.~M.,  2011, \mn@doi [\mnras]
  {10.1111/j.1365-2966.2010.17882.x}, \href
  {http://adsabs.harvard.edu/abs/2011MNRAS.411.2750S} {411, 2750}

\bibitem[\protect\citeauthoryear{{Staguhn} et~al.,}{{Staguhn}
  et~al.}{2008}]{Staguhn+etal_2008}
{Staguhn} J.,  et~al., 2008, \mn@doi [Journal of Low Temperature Physics]
  {10.1007/s10909-008-9733-6}, \href
  {http://adsabs.harvard.edu/abs/2008JLTP..151..709S} {151, 709}

\bibitem[\protect\citeauthoryear{{Staguhn} et~al.,}{{Staguhn}
  et~al.}{2014}]{Staguhn+etal_2014}
{Staguhn} J.~G.,  et~al., 2014, \mn@doi [\apj] {10.1088/0004-637X/790/1/77},
  \href {http://adsabs.harvard.edu/abs/2014ApJ...790...77S} {790, 77}

\bibitem[\protect\citeauthoryear{{Tibbs} et~al.,}{{Tibbs}
  et~al.}{2010}]{Tibbs+etal_2010}
{Tibbs} C.~T.,  et~al., 2010, \mn@doi [\mnras]
  {10.1111/j.1365-2966.2009.16023.x}, \href
  {http://adsabs.harvard.edu/abs/2010MNRAS.402.1969T} {402, 1969}

\bibitem[\protect\citeauthoryear{{Tibbs} et~al.,}{{Tibbs}
  et~al.}{2011}]{Tibbs+etal_2011}
{Tibbs} C.~T.,  et~al., 2011, \mn@doi [\mnras]
  {10.1111/j.1365-2966.2011.19605.x}, \href
  {http://adsabs.harvard.edu/abs/2011MNRAS.418.1889T} {418, 1889}

\bibitem[\protect\citeauthoryear{{Tibbs} et~al.,}{{Tibbs}
  et~al.}{2012a}]{Tibbs+etal_2012}
{Tibbs} C.~T.,  et~al., 2012a, \mn@doi [\apj] {10.1088/0004-637X/754/2/94},
  \href {http://adsabs.harvard.edu/abs/2012ApJ...754...94T} {754, 94}

\bibitem[\protect\citeauthoryear{{Tibbs}, {Paladini}  \& {Dickinson}}{{Tibbs}
  et~al.}{2012b}]{Tibbs+Paladini+Dickinson_2012}
{Tibbs} C.~T.,  {Paladini} R.,   {Dickinson} C.,  2012b, \mn@doi [Advances in
  Astronomy] {10.1155/2012/124931}, \href
  {http://adsabs.harvard.edu/abs/2012AdAst2012E..41T} {2012}

\bibitem[\protect\citeauthoryear{{Todorovi{\'c}} et~al.,}{{Todorovi{\'c}}
  et~al.}{2010}]{Todorovic+etal_2010}
{Todorovi{\'c}} M.,  et~al., 2010, \mn@doi [\mnras]
  {10.1111/j.1365-2966.2010.16809.x}, \href
  {http://adsabs.harvard.edu/abs/2010MNRAS.406.1629T} {406, 1629}

\bibitem[\protect\citeauthoryear{{Vidal} et~al.,}{{Vidal}
  et~al.}{2011}]{Vidal+etal_2011}
{Vidal} M.,  et~al., 2011, \mn@doi [\mnras] {10.1111/j.1365-2966.2011.18562.x},
  \href {http://adsabs.harvard.edu/abs/2011MNRAS.414.2424V} {414, 2424}

\bibitem[\protect\citeauthoryear{{Vilardell}, {Ribas}, {Jordi}, {Fitzpatrick}
  \& {Guinan}}{{Vilardell} et~al.}{2010}]{Vilardell+etal_2010}
{Vilardell} F.,  {Ribas} I.,  {Jordi} C.,  {Fitzpatrick} E.~L.,   {Guinan}
  E.~F.,  2010, \mn@doi [\aap] {10.1051/0004-6361/200913299}, \href
  {http://adsabs.harvard.edu/abs/2010A%26A...509A..70V} {509, A70}

\bibitem[\protect\citeauthoryear{{Ysard} \& {Verstraete}}{{Ysard} \&
  {Verstraete}}{2010}]{Ysard+Verstraete_2010}
{Ysard} N.,  {Verstraete} L.,  2010, \mn@doi [\aap]
  {10.1051/0004-6361/200912708}, \href
  {http://adsabs.harvard.edu/abs/2010A%26A...509A..12Y} {509, A12}

\bibitem[\protect\citeauthoryear{{Ysard}, {Miville-Desch{\^e}nes}  \&
  {Verstraete}}{{Ysard} et~al.}{2010}]{Ysard+Miville-Deschenes+Verstraete_2010}
{Ysard} N.,  {Miville-Desch{\^e}nes} M.~A.,   {Verstraete} L.,  2010, \mn@doi
  [\aap] {10.1051/0004-6361/200912715}, \href
  {http://adsabs.harvard.edu/abs/2010A%26A...509L...1Y} {509, L1}

\bibitem[\protect\citeauthoryear{{de Oliveira-Costa}, {Kogut}, {Devlin},
  {Netterfield}, {Page}  \& {Wollack}}{{de Oliveira-Costa}
  et~al.}{1997}]{deOliveiraCosta+etal_1997}
{de Oliveira-Costa} A.,  {Kogut} A.,  {Devlin} M.~J.,  {Netterfield} C.~B.,
  {Page} L.~A.,   {Wollack} E.~J.,  1997, \mn@doi [\apjl] {10.1086/310684},
  \href {http://adsabs.harvard.edu/abs/1997ApJ...482L..17D} {482, L17}

\makeatother
\end{thebibliography}
\label{lastpage}
\end{document}